\documentclass[english]{ieej-e}
\usepackage[dvipdfmx]{graphicx}
\usepackage[dvipdfmx]{color}
\usepackage[fleqn]{amsmath}
\usepackage[varg]{txfonts}

\usepackage{subfigure}
\usepackage{multirow}
\usepackage{cite}
\usepackage{color}

\FIELD{C}
\YEAR{2021}
\NO{2}
\title{Application of Reversible Data Hiding for Printing with Special Color Inks to Preserve Compatibility with Normal Printing}
\authorlist{%
 \authorentry{Kotoko Hiraoka}{n}{A}
 \authorentry{Kensuke Fukumoto}{n}{A}
 \authorentry{Takashi Yamazoe}{n}{B}
 \authorentry{Norimichi Tsumura}{n}{C}
 \authorentry{Satoshi Kaneko}{n}{D}
 \authorentry{Wataru Arai}{n}{D}
 \authorentry[imaizumi@chiba-u.jp]{Shoko Imaizumi}{m}{C}
}
\affiliate[A]{Graduate School of Science and Engineering, Chiba University\\
              1-33 Yayoicho, Inage-ku, Chiba-shi, Chiba 263-8522, Japan}
\affiliate[B]{Institute for Global Prominent Research, Chiba University\\
              1-33 Yayoicho, Inage-ku, Chiba-shi, Chiba 263-8522, Japan}
\affiliate[C]{Graduate School of Engineering, Chiba University\\
              1-33 Yayoicho, Inage-ku, Chiba-shi, Chiba 263-8522, Japan}
\affiliate[D]{Software Design Department, MIMAKI ENGINEERING CO., LTD.\\
              2182-3 Shigeno-otsu, Tomi-shi, Nagano 389-0512, Japan}

\received{2020}{4}{3}
\revised{2020}{8}{17}

\begin{document}
\begin{abstract}
We propose an efficient framework with compatibility between normal printing and printing with special color inks in this paper.
Special color inks can be used for printing to represent some particular colors and specific optical properties, which are difficult to express using only CMYK inks.
Special color layers are required in addition to the general color layer for printing with special color inks.
We introduce a reversible data hiding (RDH) method to embed the special color layers into the general color layer without visible artifacts. 
The proposed method can realize both normal printing and printing with special color inks by using a single layer. 
Our experimental results show that the quality of the marked image is virtually identical to that of the original image, i.e., the general color layer.
\end{abstract}
\maketitle

\section{Introduction}
Normal printing is generally based on subtractive color mixing using cyan (C), magenta (M), yellow (Y), and black (K) inks, which are used for color representation.
The image data for normal printing is contained in the general color layer, which originally consists of RGB components, and is converted to CMYK color space on printing.
On printing using CMYK inks, the colors that can be reproduced are in the color gamut associated with the printing process.
Therefore, special color inks are required to represent colors such as vivid orange, violet, and green.
In particular, when the printing surface is transparency or colored, white ink is applied to represent white color.
These special color inks have been previously prepared for specific colors that are difficult to represent using only CMYK inks and can be used to expand the range of color representation.
In addition to inks that represent specific colors, there also exist some special color inks that can represent specific optical properties in an image, e.g., gloss and fluorescence.
For instance, printing with silver ink, called metallic color printing, can represent the gloss on a metal surface.
This gloss cannot be represented by mixing normal color inks.

Data hiding techniques have been widely studied in the information security field for protecting the copyright of digital images~\cite{Stegano, Analysis, Interpolation, Robust, Survey, Ni, Contrast, Tian, Auto, Map, Spatial, Frequency}.
The main purpose of data hiding is to prohibit malicious copying and secondary use without permission by embedding the copyright information into the images.
Its performance is primarily discussed in terms of the quality of the marked image, data-hiding capacity, and robustness against some attacks.
Data hiding is divided into two categories: irreversible data hiding (IDH) and reversible data hiding (RDH)~\cite{Ni, Contrast, Tian, Auto, Map, Spatial, Frequency}.
The former somehow deteriorates quality of a retrieved image even when payload is extracted, whereas the latter can perfectly restore the original image after extracting the payload.
IDH methods can generally embed a large amount of payload without visible artifacts, and is also robust to attacks, such as clipping, rotating, and scaling.
In contrast, RDH methods embed less amount of payload and are less robust to such attacks compared to IDH methods, while they can retrieve the intact image.
One RDH method, e.g., the Haar discrete wavelet transformation (HDWT)-based method, first transforms an original image into the HDWT-based frequency domain and then embeds payload into high-frequency coefficients~\cite{Frequency}.
The quality of the marked image can be relatively high in this method.
Another method uses difference expansion~\cite{Tian}. 
This method is based on expanding differences between each pair of neighboring pixels to embed a large amount of payload. 

In regards to printed matters, the multiple techniques of data hiding have been proposed~\cite{printing1, printing2, printing3}.
In these literatures, payload is first embedded into the image data, and then the marked image is printed. 
After returning the printed matter to the electronic data by a scanner or camera, the payload can be extracted precisely. 
Those methods have resilience against the print-scan process but cannot restore the original electronic data.  
To our best knowledge, there is no technique that deals with printing with special color inks to consider the compatibility with normal printing. 
Although our proposed method does not consider the print-scan process, it can completely restore the original layers in electronic form.

\begin{figure}[t]
\centering
\includegraphics[scale=0.275]{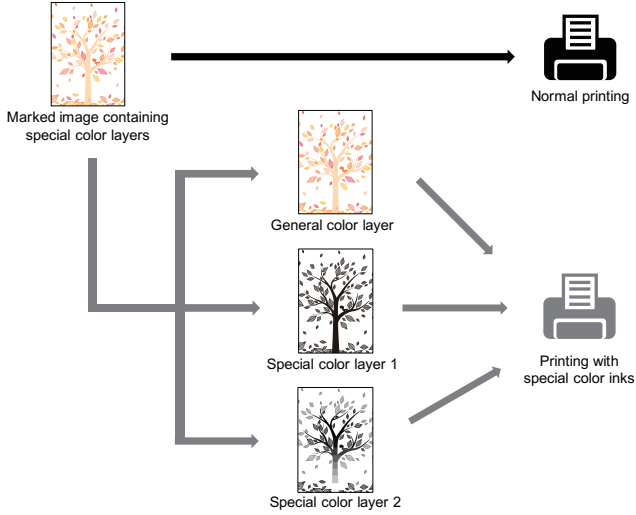}
\caption{Strategy to preserve compatibility between normal printing and printing with special color inks.}
\label{fig.1}
\end{figure}

\begin{figure}[t]
\centering
\includegraphics[scale=0.40]{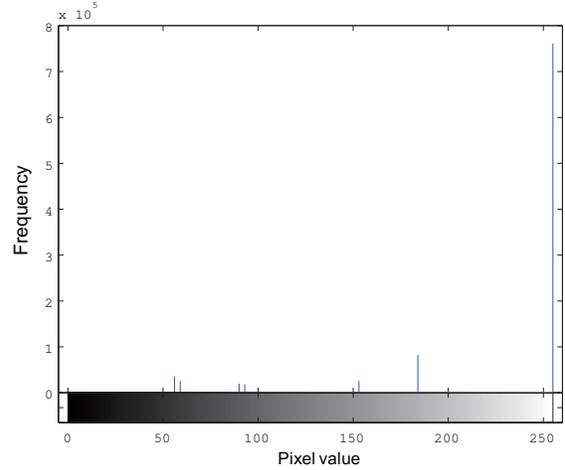}
\caption{Histogram of illustration depicted in Fig.~\ref{fig.1}.}
\label{fig.2}
\end{figure}

In this paper, we propose a new framework for preserving compatibility between normal printing and printing with special color inks. 
This framework embeds the special color layers into the general color layer without visible artifacts. 
We show the utilization of the marked image containing special color layers in Fig.~\ref{fig.1}.
In normal printing, the marked color-layer, i.e., the marked image can be directly sent to the printer.
In contrast, on printing with special color inks, the special color layers are reversibly extracted from the marked image and the general color layer is also retrieved perfectly.
Consequently, we can suppress the total amount of data by using this framework.
The target images in this paper are illustrations. 
They are widely used for promotional posters, postcards, and so forth, where special color inks may be regionally applied for eye-catching.
An illustration is frequently represented with a limited number of colors; thus, the histogram is scattered as shown in Fig.~\ref{fig.2}. 
We focus on such sparse histogram images, and high efficacy can be attained for them.
We evaluate the quality of the marked images through our experiment and confirm that the marked images produced by the proposed method can achieve high quality without affecting the normal printing.

\section{Preliminary}
\subsection{Printing with special color inks}
\begin{figure}[t]
\begin{center}
\includegraphics[width=0.3\linewidth]{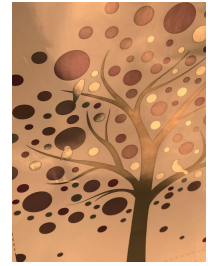}
\caption{Printed matter with silver ink.}
\label{fig.3}
\end{center}
\end{figure}

Figure~\ref{fig.3} indicates a photograph, which was printed with silver ink. 
To perform the printing with special color inks, it is necessary to provide two kinds of image data: general color layer and special color layer.
The special color layer is image data, which is often obtained as a multivalued image, and specifies the area and ink density considering the design of the printing.
In the density of the special color ink, 0 \% indicates no printing, while 100 \% means full printing.
In this paper, we assume that there are two types of special color layers: binary layer and 3-bit layer. 

\begin{figure*}[t]
\centering
\subfigure[Original image] {
\includegraphics[scale=0.33]{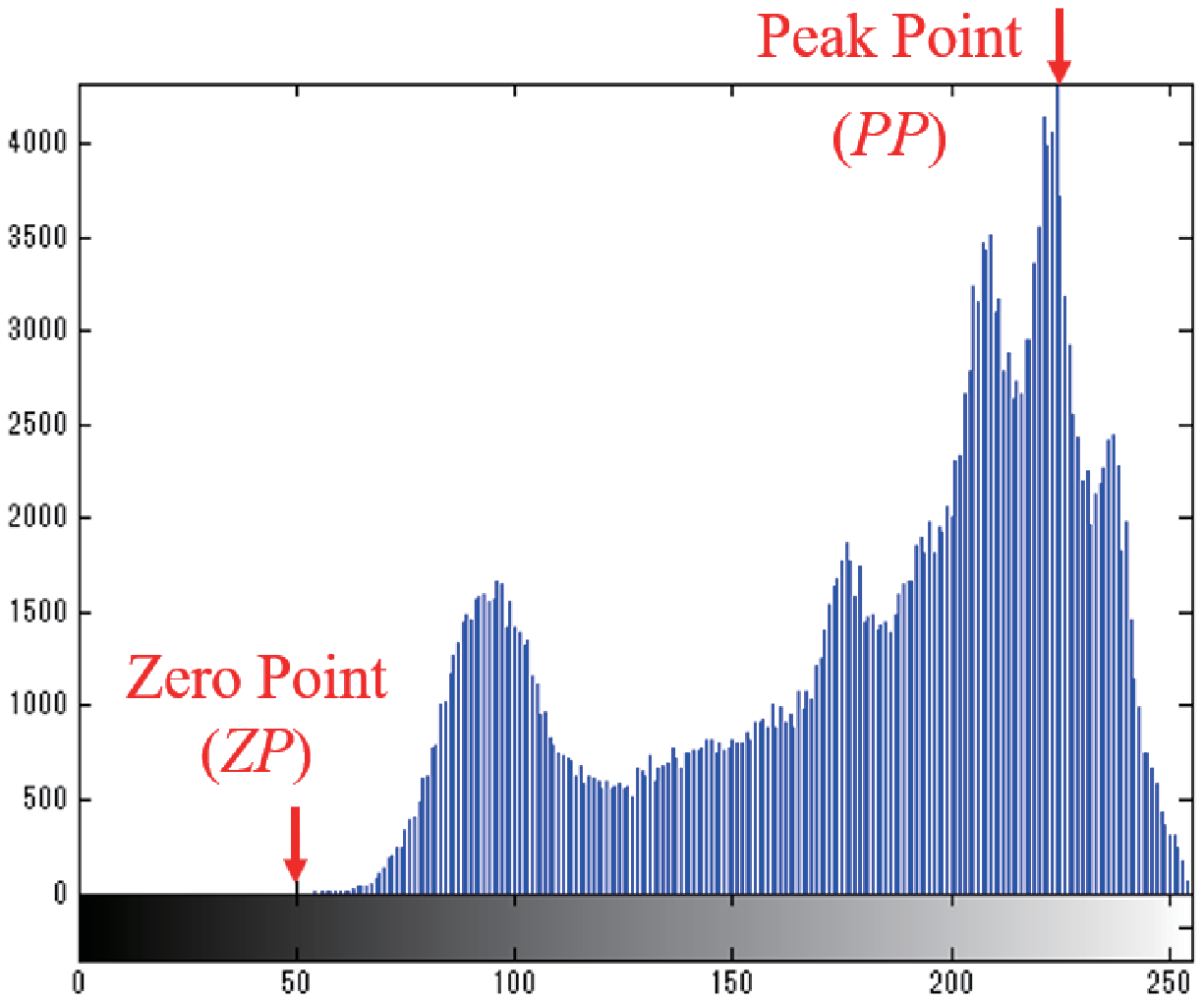}
\label{fig.4a}
} 
\subfigure[Histogram shift] {
\includegraphics[scale=0.33]{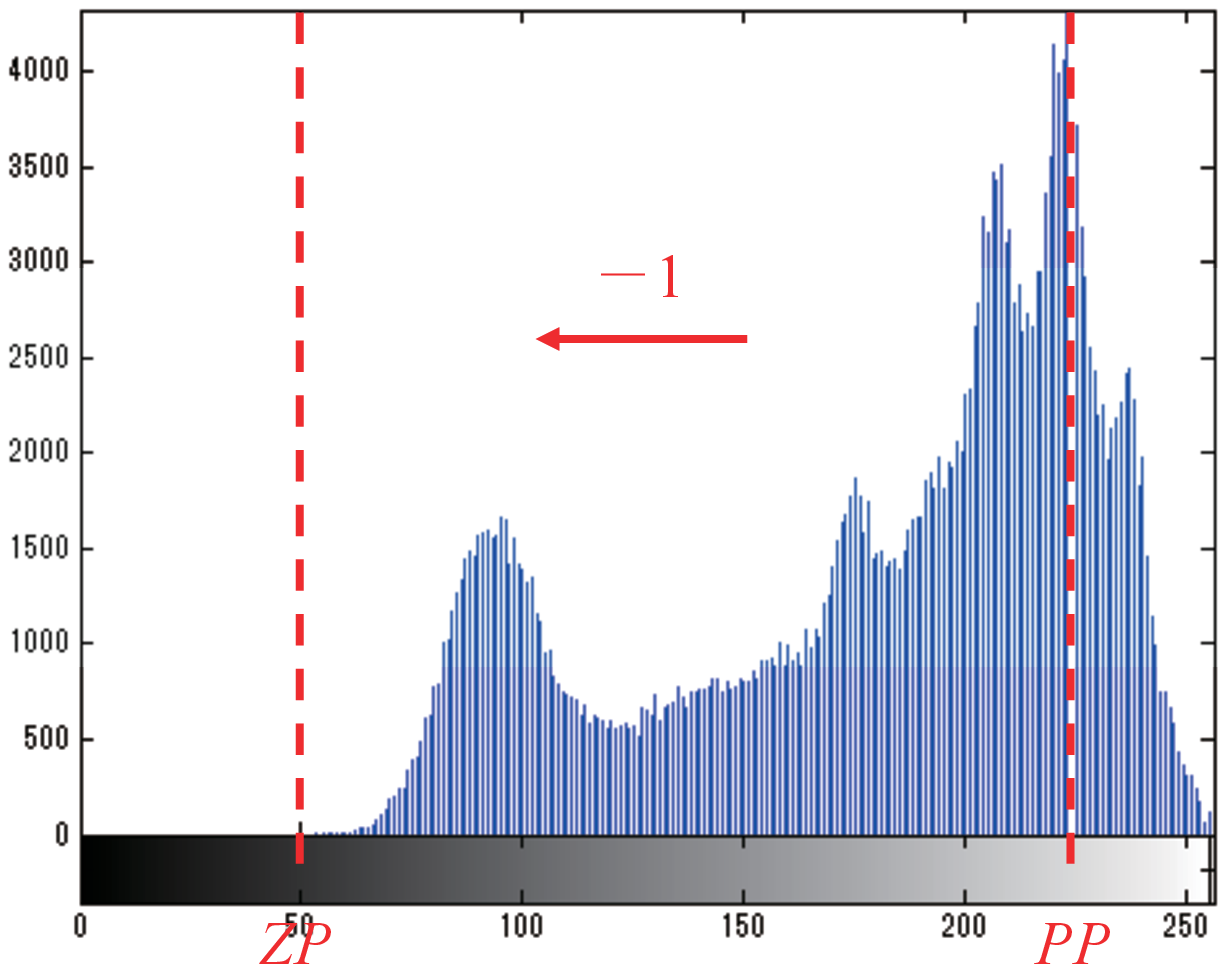}
\label{fig.4b}
}
\subfigure[Marked image] { 
\includegraphics[scale=0.33]{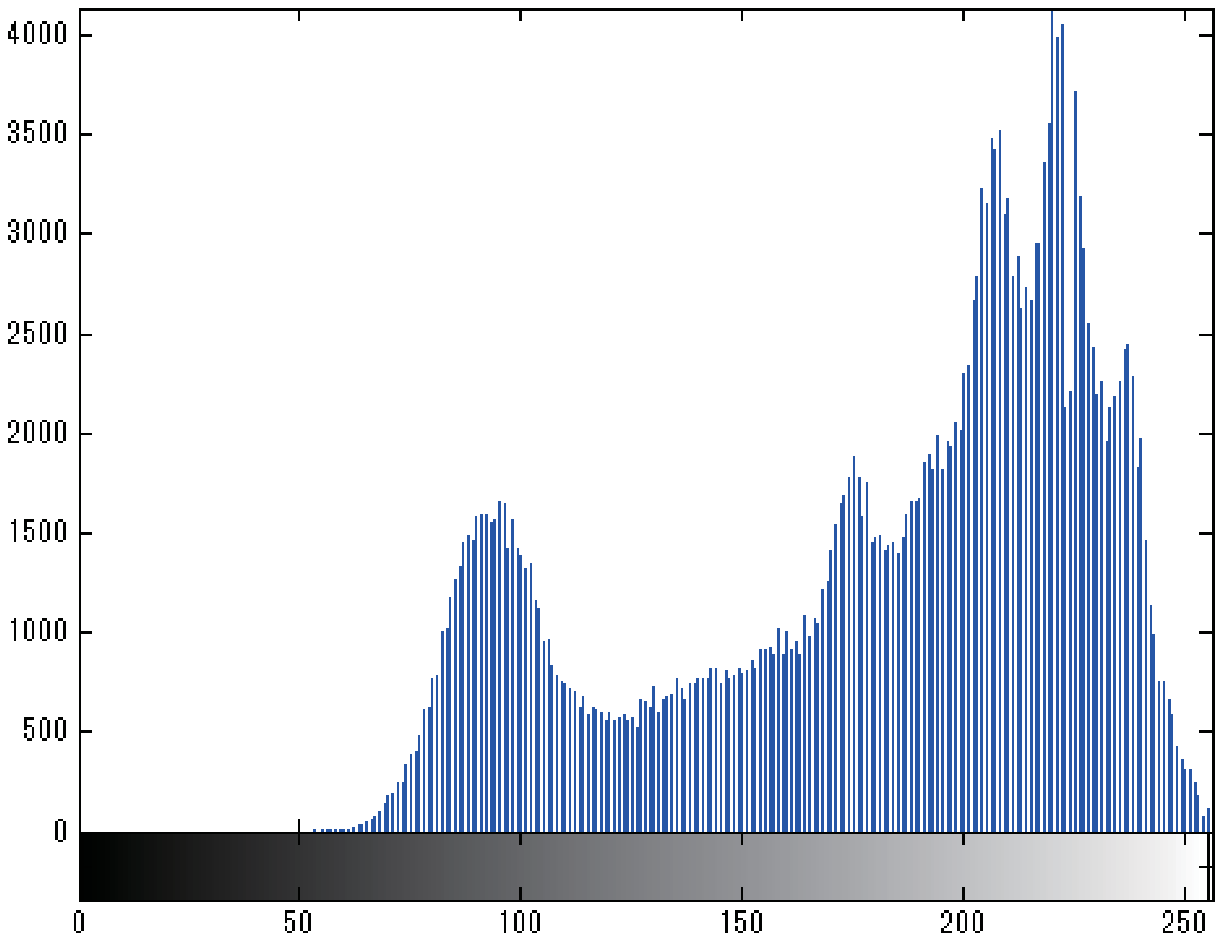}
\label{fig.4c}
}
\caption{Histograms by HS process.}
\label{fig.4}
\end{figure*}

\begin{figure*}[t]
\centering
\subfigure[Data hiding procedure]{
\includegraphics[width=0.48\linewidth]{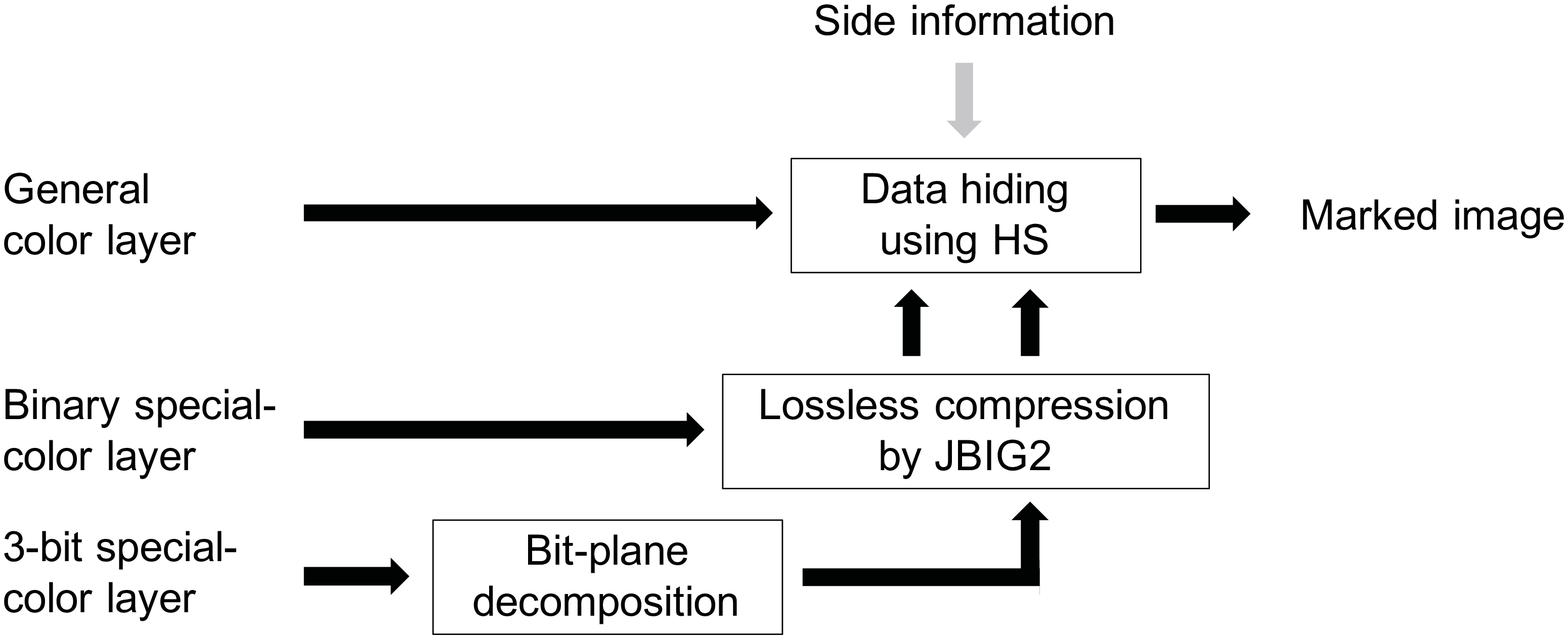}
\label{fig.5a}
}
\subfigure[Data extraction procedure]{
\includegraphics[width=0.48\linewidth]{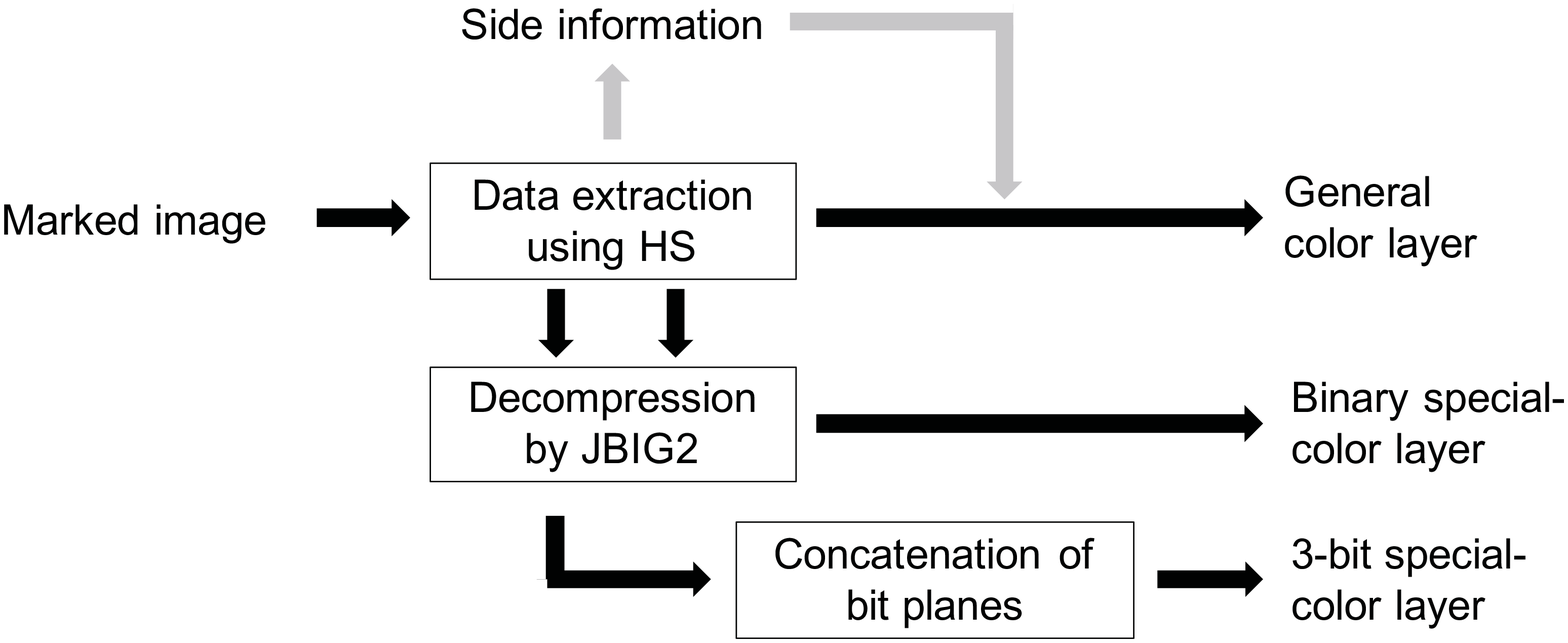}
\label{fig.5b}
}
\caption{Block diagram of proposed method.}
\label{fig.5}
\end{figure*}

\begin{figure*}[t]
\begin{center}
\subfigure[3-bit special-color layer] {
\includegraphics[width=0.15\linewidth]{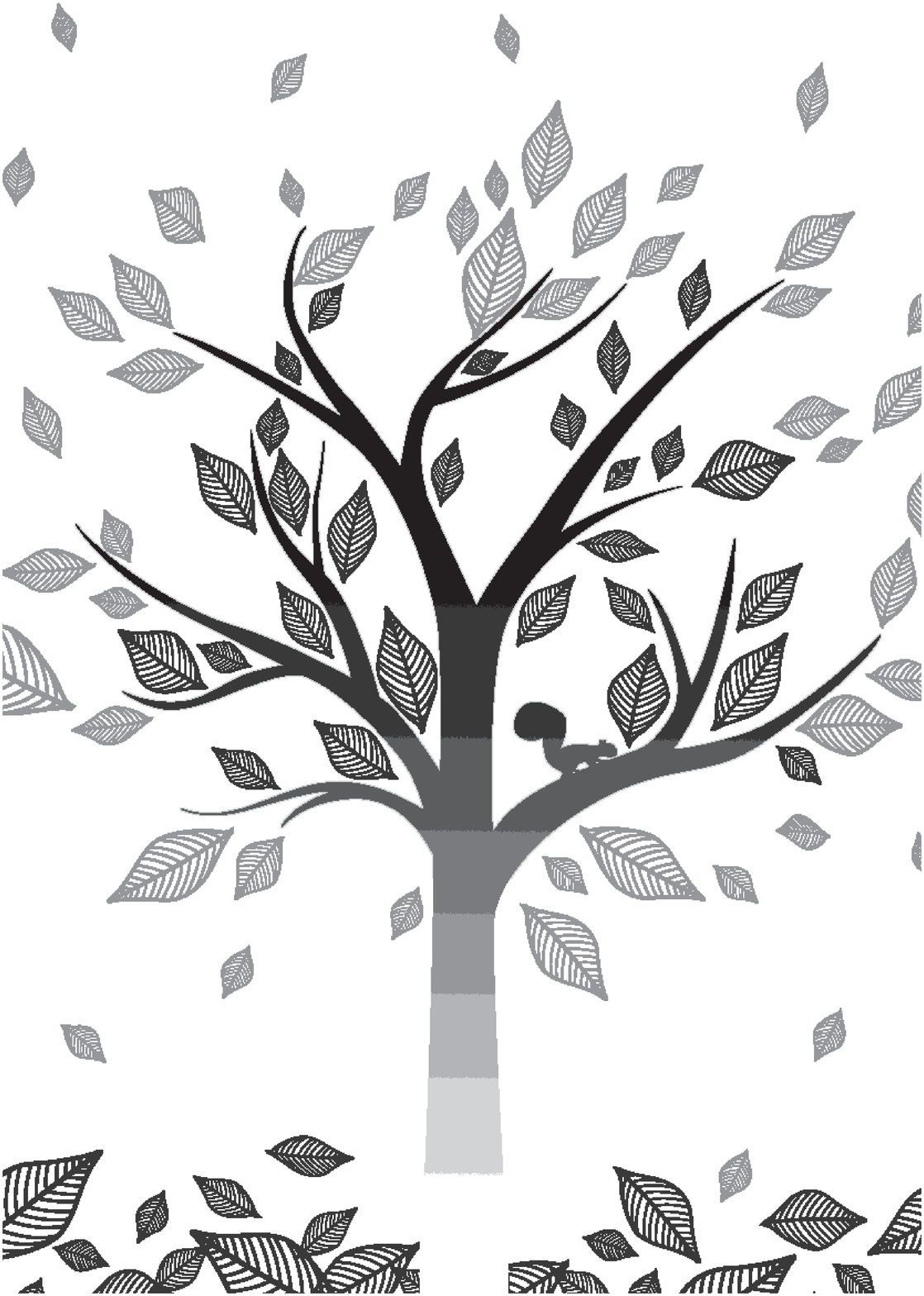}
\label{fig.6a}
}
\subfigure[Concatenation of three bit planes] {
\includegraphics[width=0.45\linewidth]{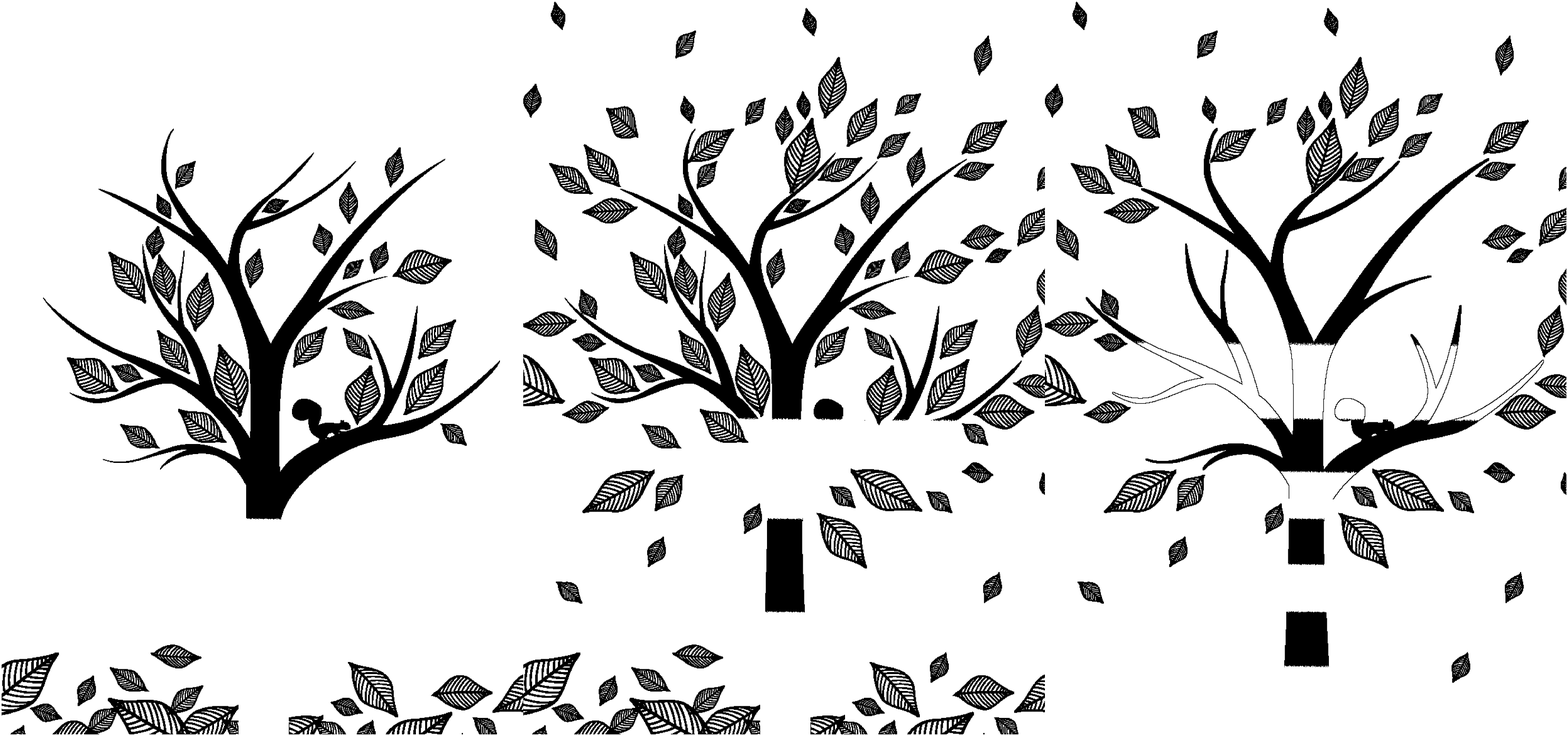}
\label{fig.6b}
}
\caption{Bit-plane decomposition and concatenation.}
\label{fig.6}
\end{center}
\end{figure*}

\subsection{Reversible data hiding based on histogram shifting~\cite{Ni}}
\label{ssec:2-2}
We introduce a histogram-based RDH method, called histogram shifting (HS) method~\cite{Ni}, in this paper. 
This method can suppress the deterioration of marked image quality by slightly changing an image histogram.
The data hiding procedure of the HS method is as follows.
Note that the bin size is 1 in this method because an original image should be perfectly retrieved after extracting payload.
Thus, the pixel values correspond to the bin values. 

\begin{description}
\item[Step 1] A pair of a peak point ($PP$) and a zero point ($ZP$) are explored from an image histogram as shown in Fig.~\ref{fig.4a}.
$PP$ and $ZP$ are the bins with the highest frequency and no pixel in the histogram, respectively.
If there is no $ZP$ in the histogram, the bin with the lowest frequency ($LP$) is adopt.
$PP$, $ZP$/$LP$, and the information of the pixels that originally possess the value of $LP$ should be stored for the extraction process.
\item[\bf{Step 2}:] All the pixels $x$ between $PP$ and $ZP$ are shifted according to the following equation.
  \begin{equation}
       x' = \left\{ \begin{array}{ll}
         x + 1, \quad x \in (PP, ZP) ~~ & {\rm if} ~~ PP<ZP \\
         x - 1, \quad x \in (ZP, PP) ~~ & {\rm if} ~~ PP>ZP.
  \end{array} \right.
\end{equation}
Accordingly, the adjacent bin of $PP$ becomes empty. 
Figure~\ref{fig.4b} shows the case of $PP > ZP$, where the bin of $PP-1$ becomes empty.
\item[Step 3] Embed payload into the pixels $x_{PP}$, where the pixel value is $PP$.
If the to-be-embedded bit is 1, the pixel is moved to the empty bin: 
\begin{equation}
      x'_{PP} = \left\{ \begin{array}{ll}
        x_{PP}+1 & {\rm if} ~~ PP<ZP \\
        x_{PP}-1 & {\rm if} ~~ PP>ZP.
      \end{array} \right.
\end{equation}
In contrast, if the to-be-embedded bit is 0, the pixel is unchanged:
\begin{equation}
x_{PP}' = x_{PP}.
\end{equation}
\end{description}
Figure~\ref{fig.4c} depicts the histogram of the marked image.
In this method, the data hiding capacity is equal to the total number of pixels $x_{PP}$.
If the number of $x_{PP}$ is less than the payload amount, these steps are repeatedly performed until all the payload is embedded.

An arbitrary data hiding technique can be used for embedding special color layers.
Nonetheless, when the image histogram is scattered, HS method is particularly effective to embed the payload with less distortion.

\begin{figure*}[t]
\centering
\subfigure[Image 1]{
\includegraphics[width=0.15\linewidth]{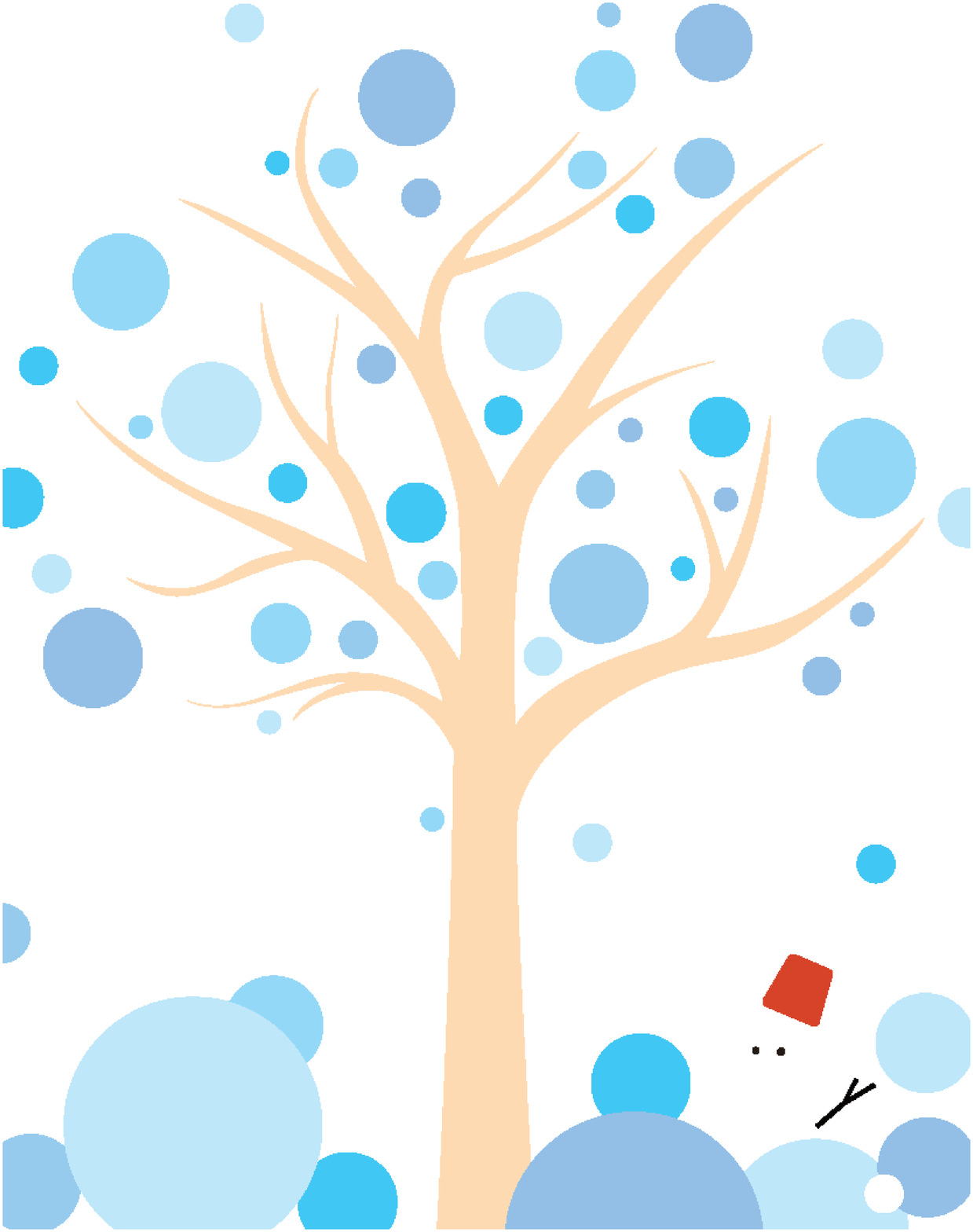}
}
\subfigure[Image 2]{
\includegraphics[width=0.15\linewidth]{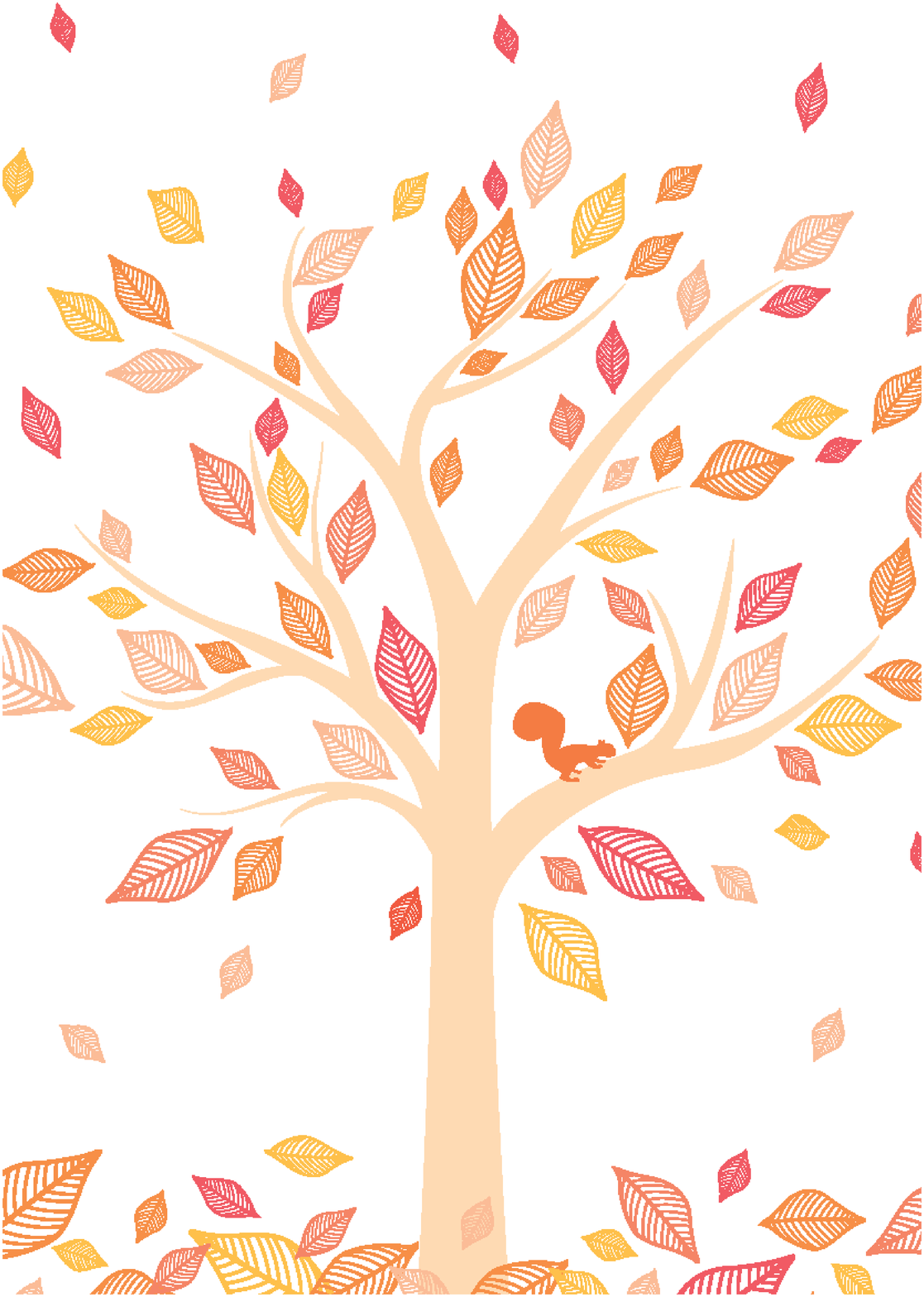}
}
\subfigure[Image 3]{
\includegraphics[width=0.15\linewidth]{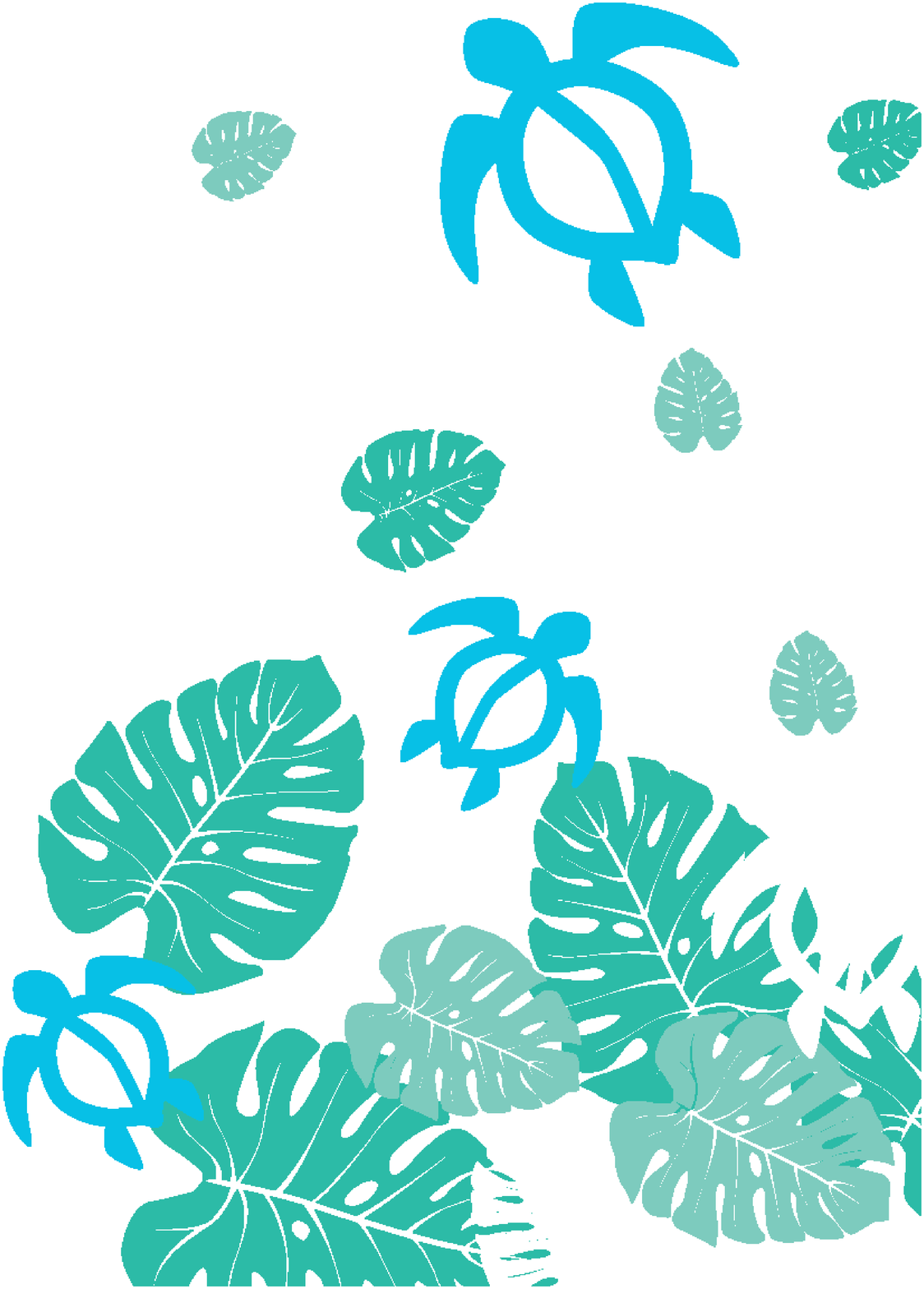}
}
\subfigure[Image 4]{
\includegraphics[width=0.15\linewidth]{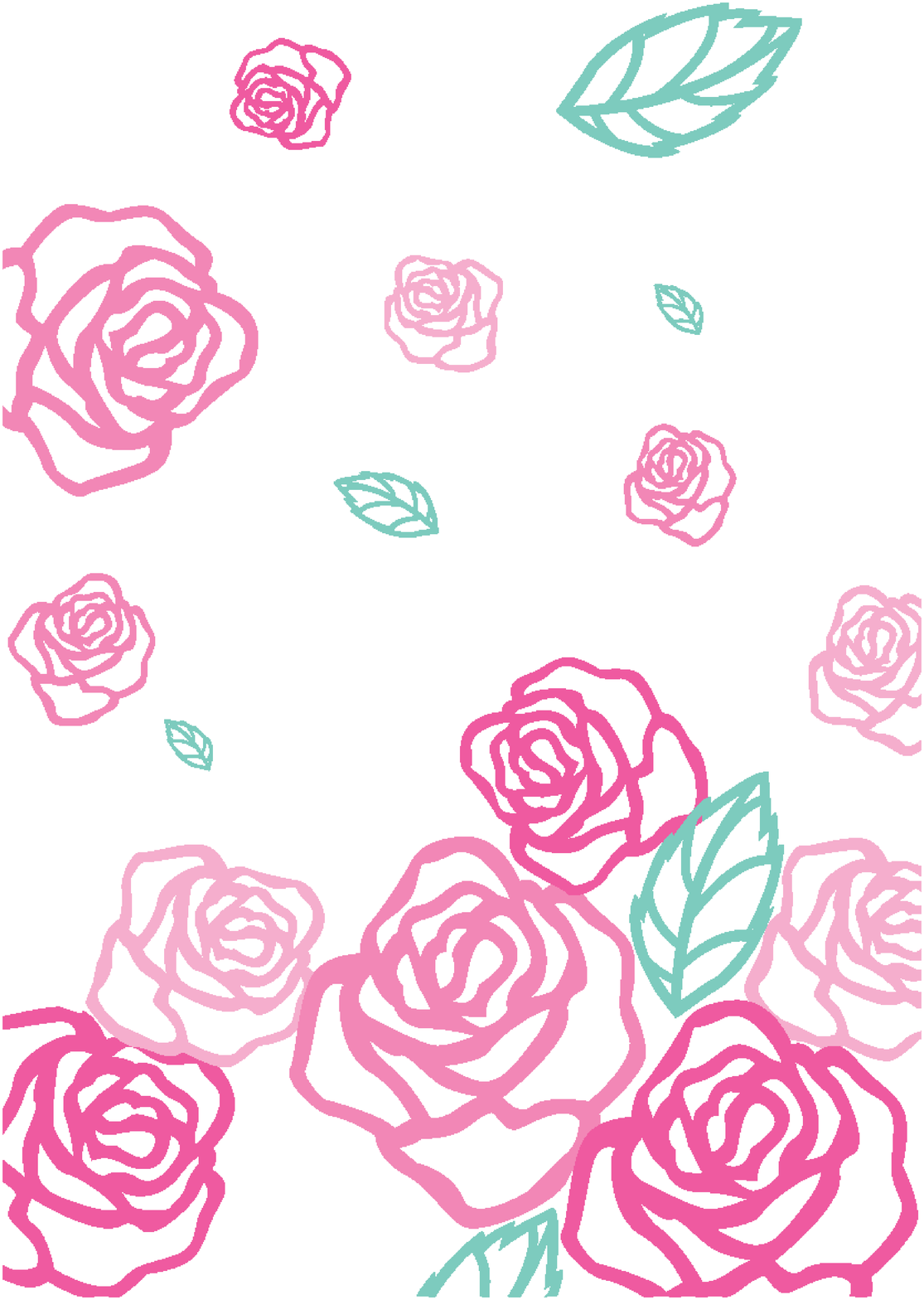}
}
\subfigure[Image 5]{
\includegraphics[width=0.12\linewidth]{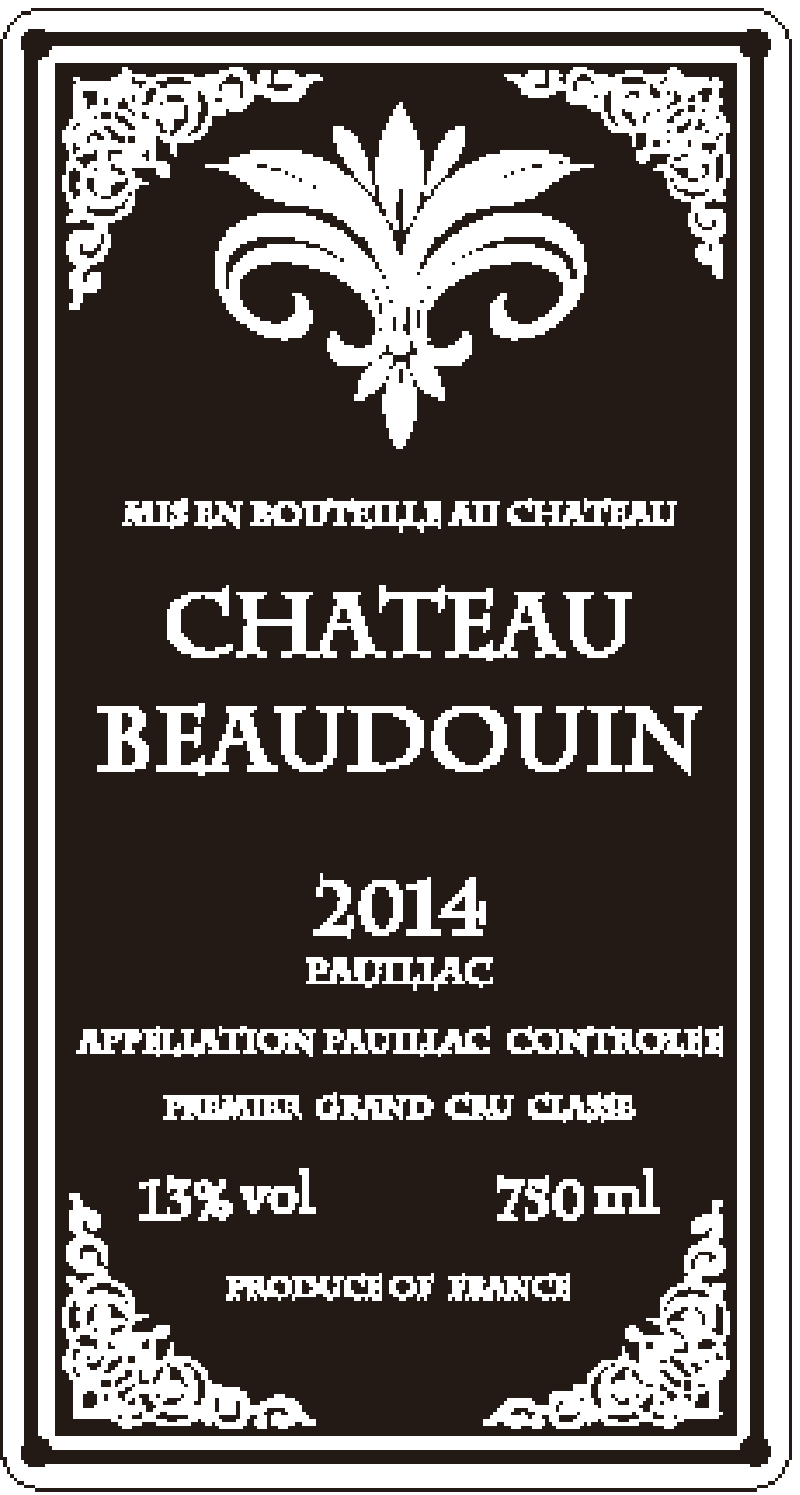}
}
\caption{Original images, i.e., general color layers (Images 1-4: 830$\times$1170 pixels, Image 5: 228$\times$428 pixels).}
\label{fig.7}
\end{figure*}

\section{Proposed method}
In this section, we propose a new framework for printing with special color inks.
The proposed method embeds the image data of special color layers, such as a silver layer and a white layer, into the general color layer without visible distortion. 
In this paper, we focus on illustrations with a limited number of colors as the target images.  
The histograms of these illustrations tend to be scattered; thus, we utilize HS method for RDH.

Here, we assume that two special color layers, which are a binary layer and a 3-bit layer, are embedded into the general color layer.  
Figure~\ref{fig.5} is the block diagram of the proposed algorithm. 
We describe the preprocessing for the special color layers, and the data hiding and extraction procedures of our method in what follows.

\subsection{Preprocessing}
The 3-bit special-color layer is decomposed into three bit planes, and they are concatenated in a horizontal direction as shown in Fig.~\ref{fig.6}.
Then, the concatenated image is compressed by JBIG2~\cite{JBIG2}, which is an international image compression standard for binary images. 
The other special color layer, which is originally binary, is also compressed by JBIG2 in advance.

\subsection{Data hiding procedure}
\label{ssec:3-2}
The compressed special-color layers are embedded into the general color layer in RGB color space by using HS method .
The detailed process is described in \ref{ssec:2-2}.
It should be noted that the order to embed payload into the pixels $x_{PP}$ is not specifically determined in any literatures about HS method.
In this paper, we explore the pixels $x_{PP}$ in the raster-scan order and embed the 1-bit payload (0 or 1) into each $x_{PP}$. 
In the same manner, the payload can be perfectly extracted.

In case that there is more than one peak point, the peak point with the smallest pixel value is adopted as $PP$.
In contrast, when there are multiple zero points, the zero point with the pixel value closest to $PP$ is set to $ZP$. 
If there is no $ZP$ in the histogram, $LP$ is adopted.
In that case, the pixels with the values of $LP$ and $LP-1/LP+1$ are merged by HS process.
The information of the pixels, which originally have the value of $LP$, should be embedded together with the pure payload.

The 8-bit values of $PP$ and $ZP$/$LP$ are also necessary in the extraction process.
It is required that they are also stored as side information.
One of the most traditional ways to embed the values of $PP$ and $ZP$/$LP$ is using the least significant bits (LSBs) \cite{Contrast}.
We embed these values into the LSBs of the first 16 pixels in the bottom row. 
The original LSBs are also embedded together with the pure payload. 

Besides, human eyes have a lower sensitivity in the order corresponding to B, R and G component.
Therefore, in our case, the 3-bit and the binary special-color layers are embedded into B and R components, respectively. 

\begin{figure*}[t]
\centering
\subfigure[Image 1]{
\includegraphics[width=0.15\linewidth]{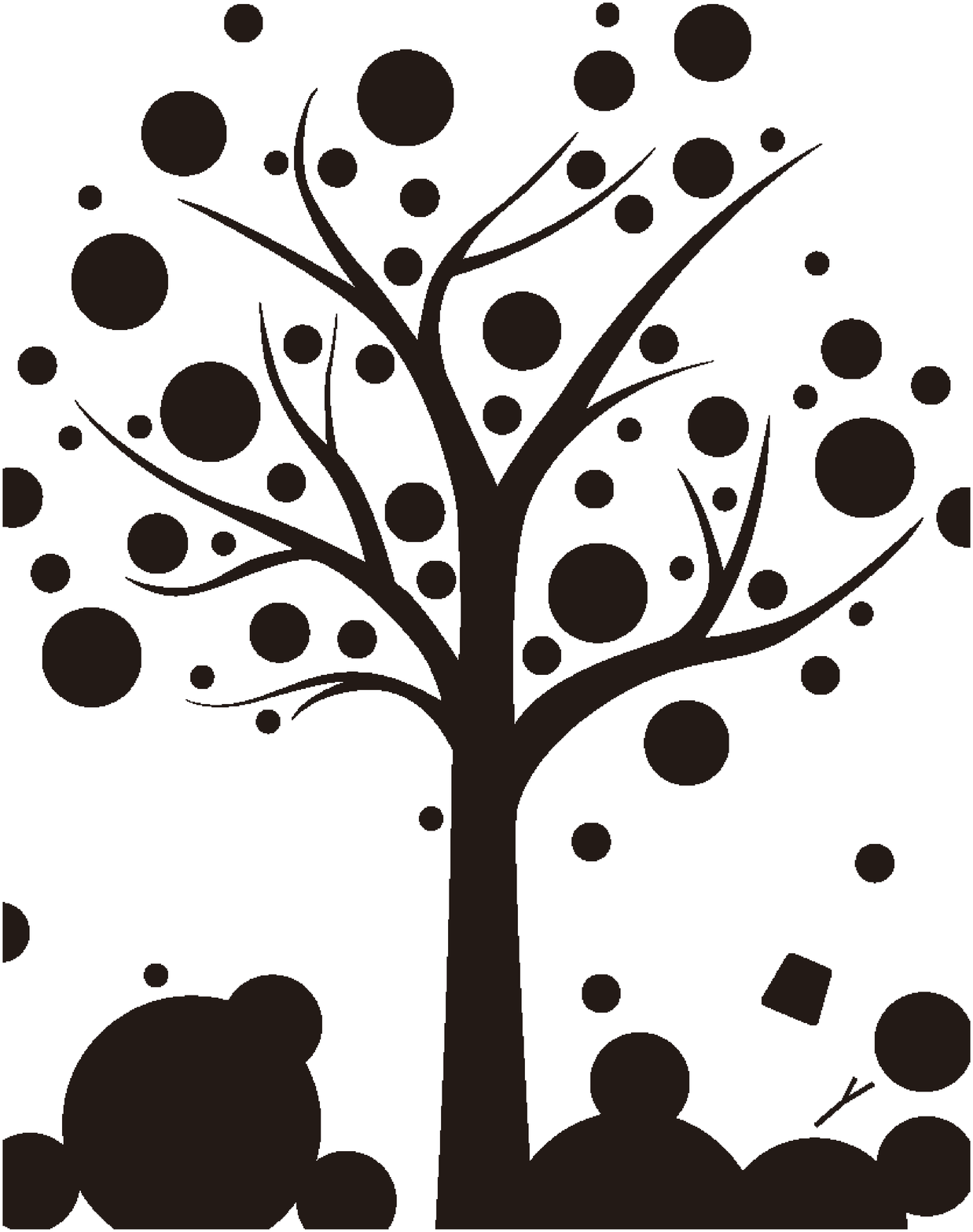}
}
\subfigure[Image 2]{
\includegraphics[width=0.15\linewidth]{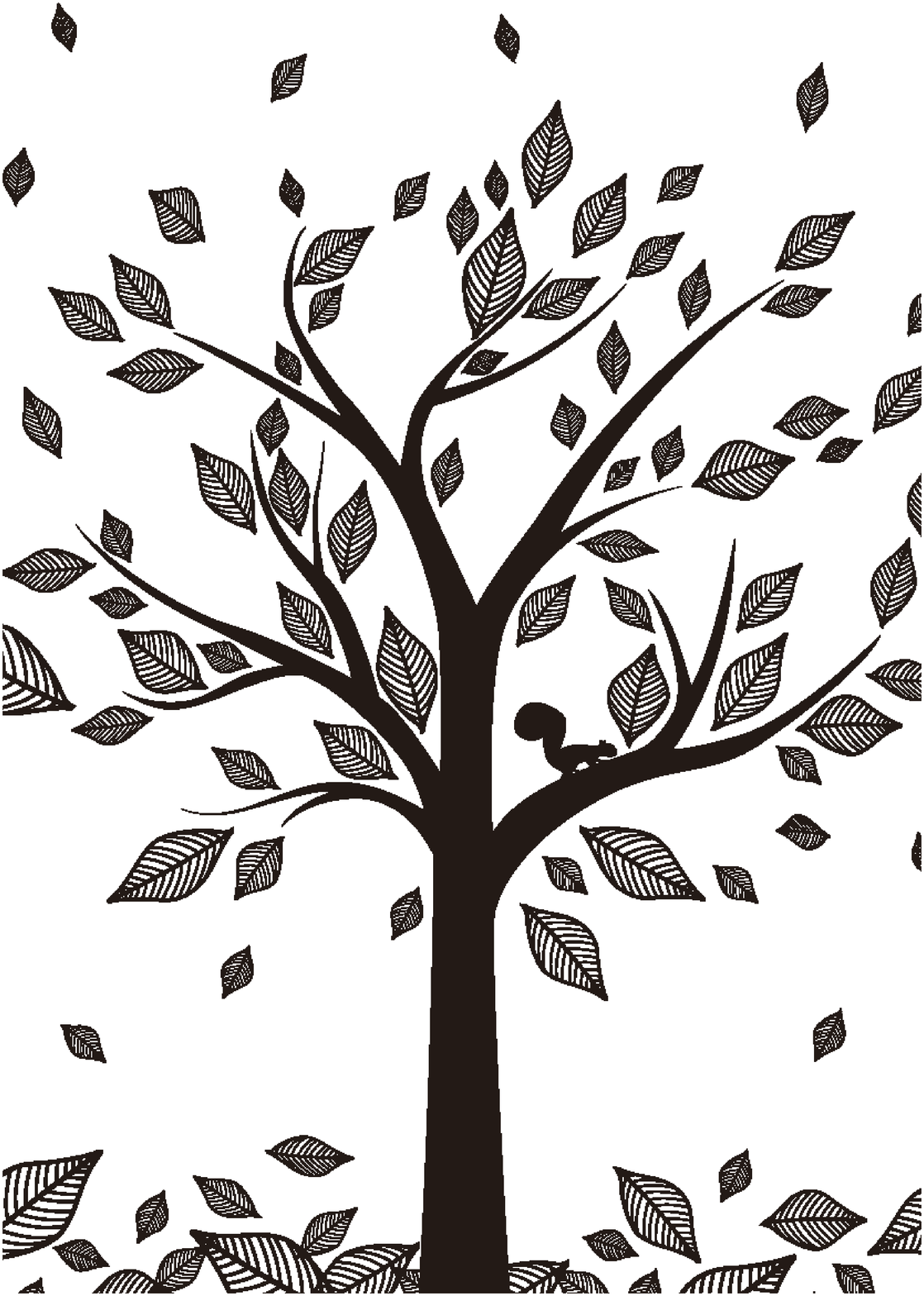}
}
\subfigure[Image 3]{
\includegraphics[width=0.15\linewidth]{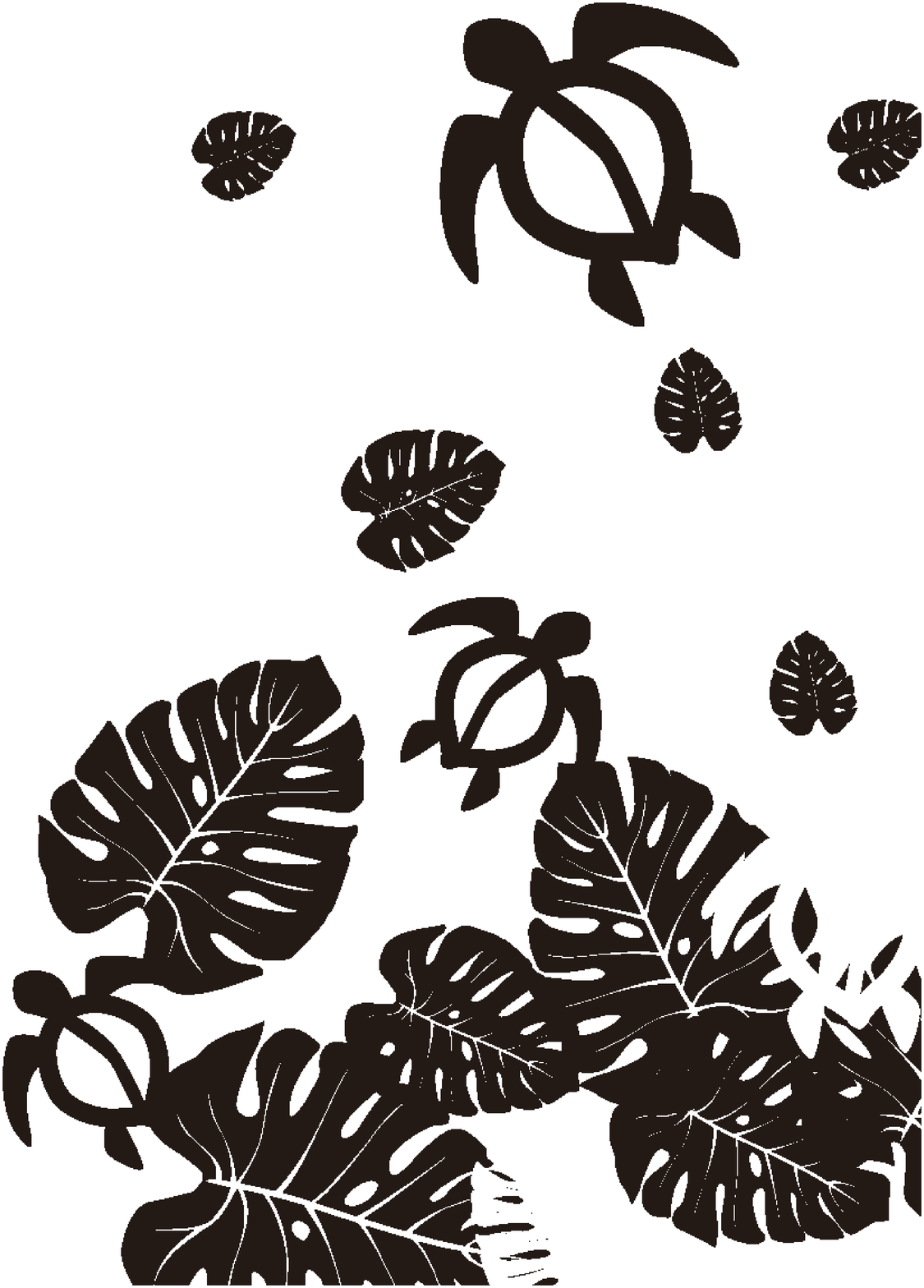}
}
\subfigure[Image 4]{
\includegraphics[width=0.15\linewidth]{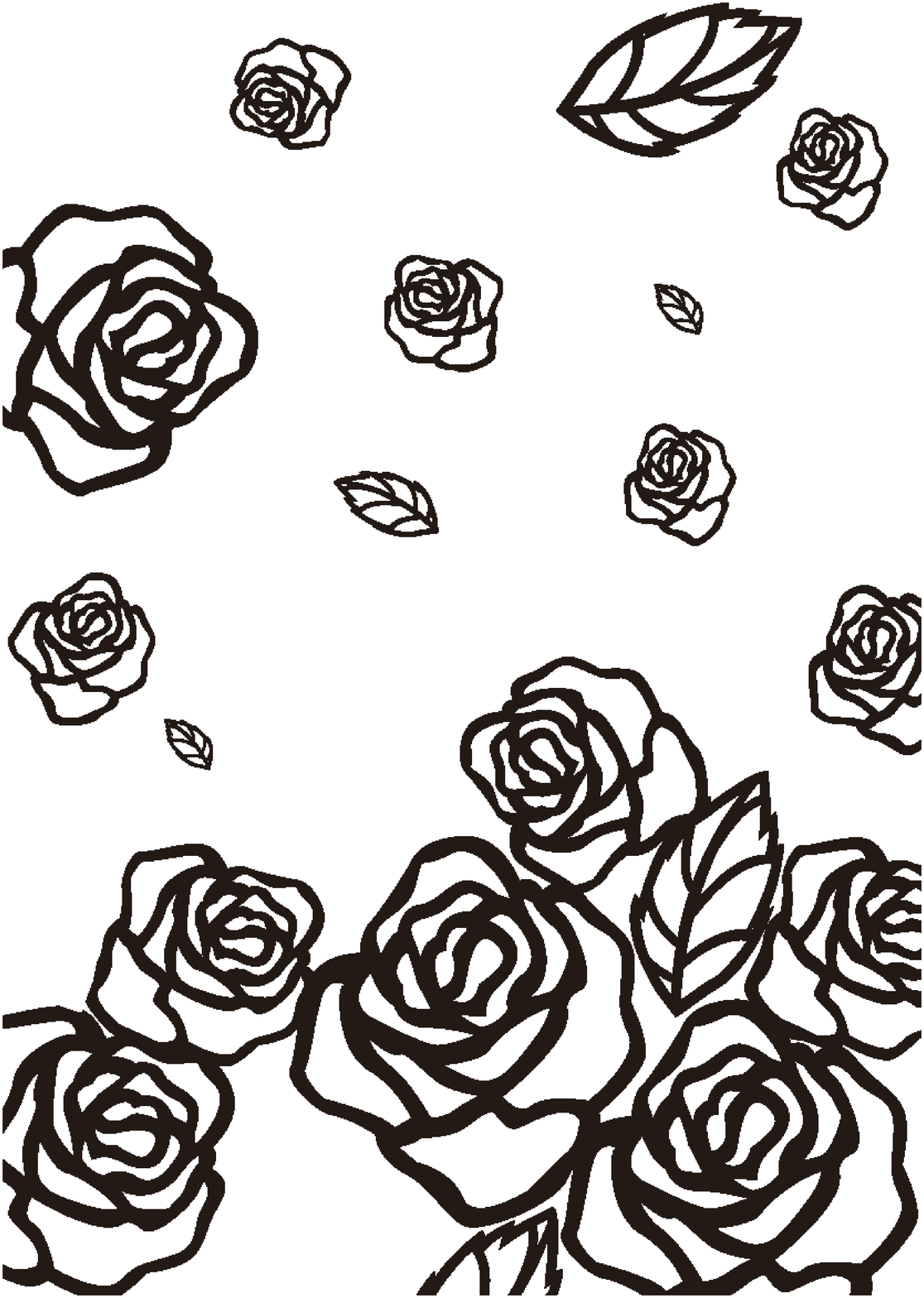}
}
\subfigure[Image 5]{
\includegraphics[width=0.12\linewidth]{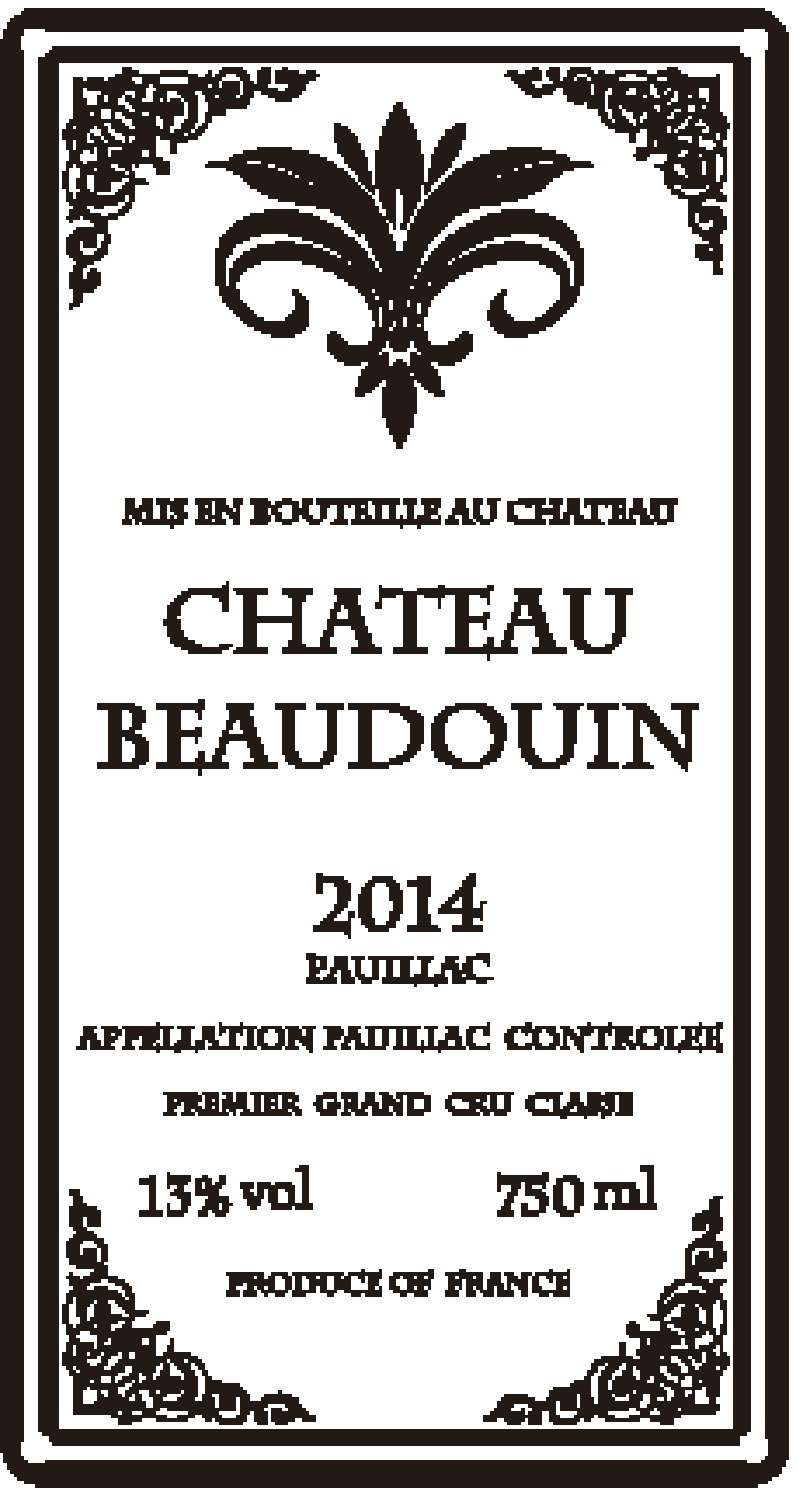}
}
\caption{Binary special-color layers.}
\label{fig.8}
\end{figure*}

\begin{figure*}[t]
\centering
\subfigure[Image 1]{
\includegraphics[width=0.45\linewidth]{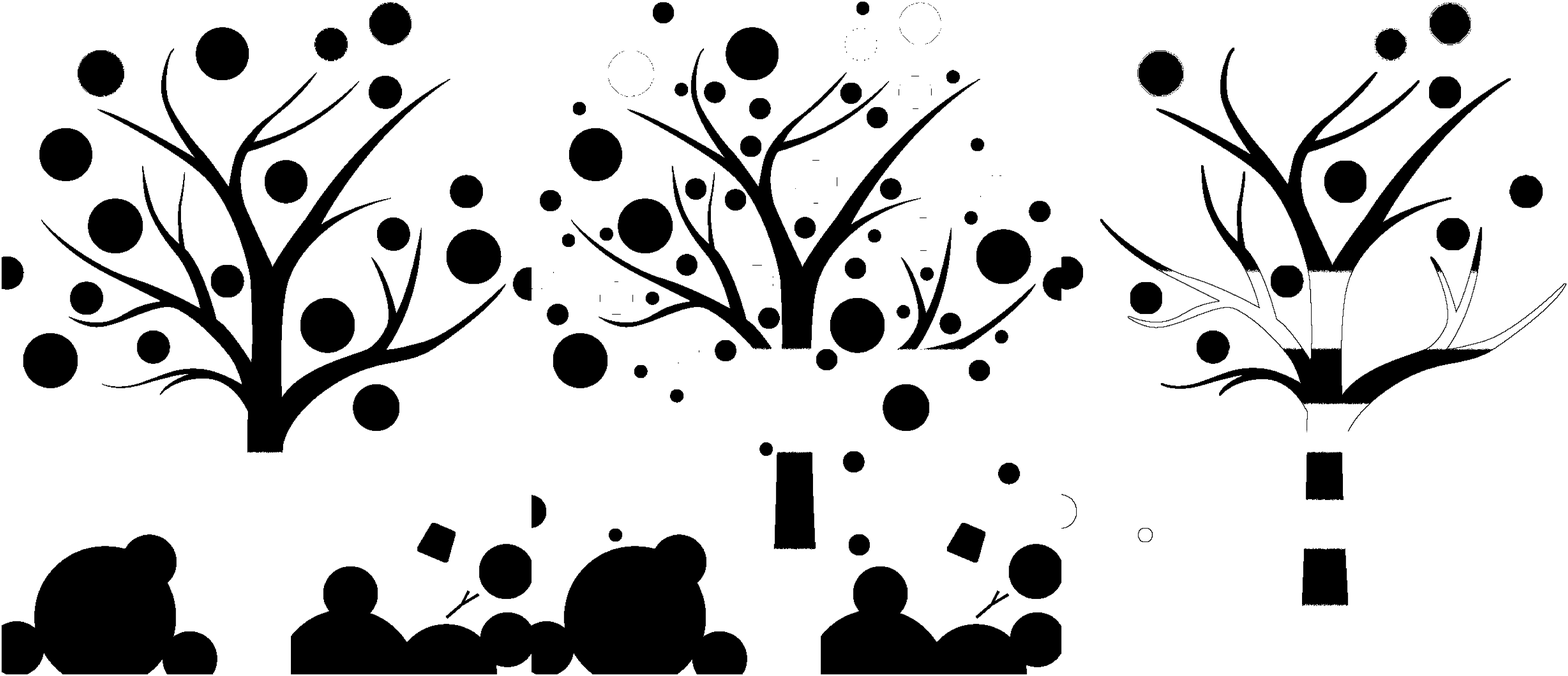}
}
\subfigure[Image 2]{
\includegraphics[width=0.45\linewidth]{img/autumn_sm3abch.eps}
} \\
\subfigure[Image 3]{
\includegraphics[width=0.45\linewidth]{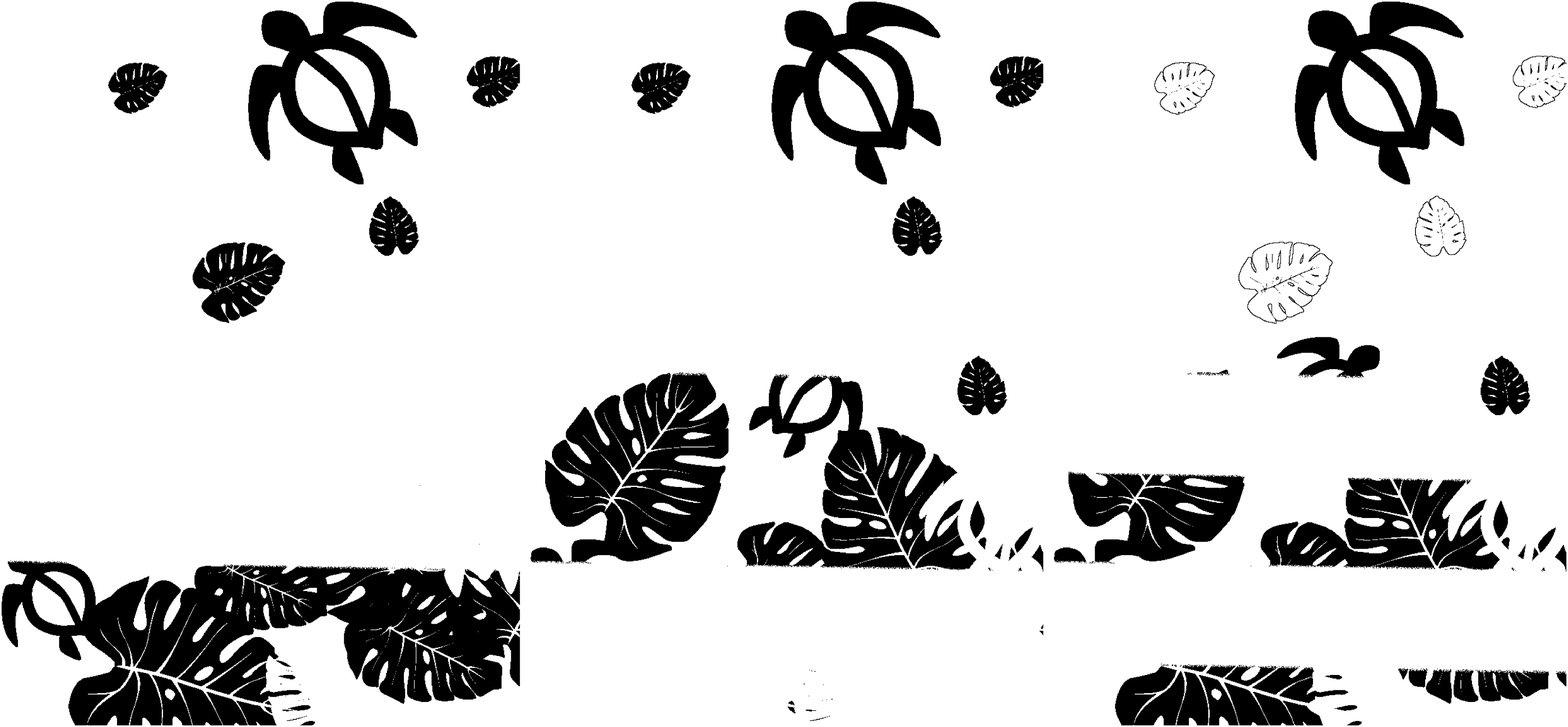}
}
\subfigure[Image 4]{
\includegraphics[width=0.45\linewidth]{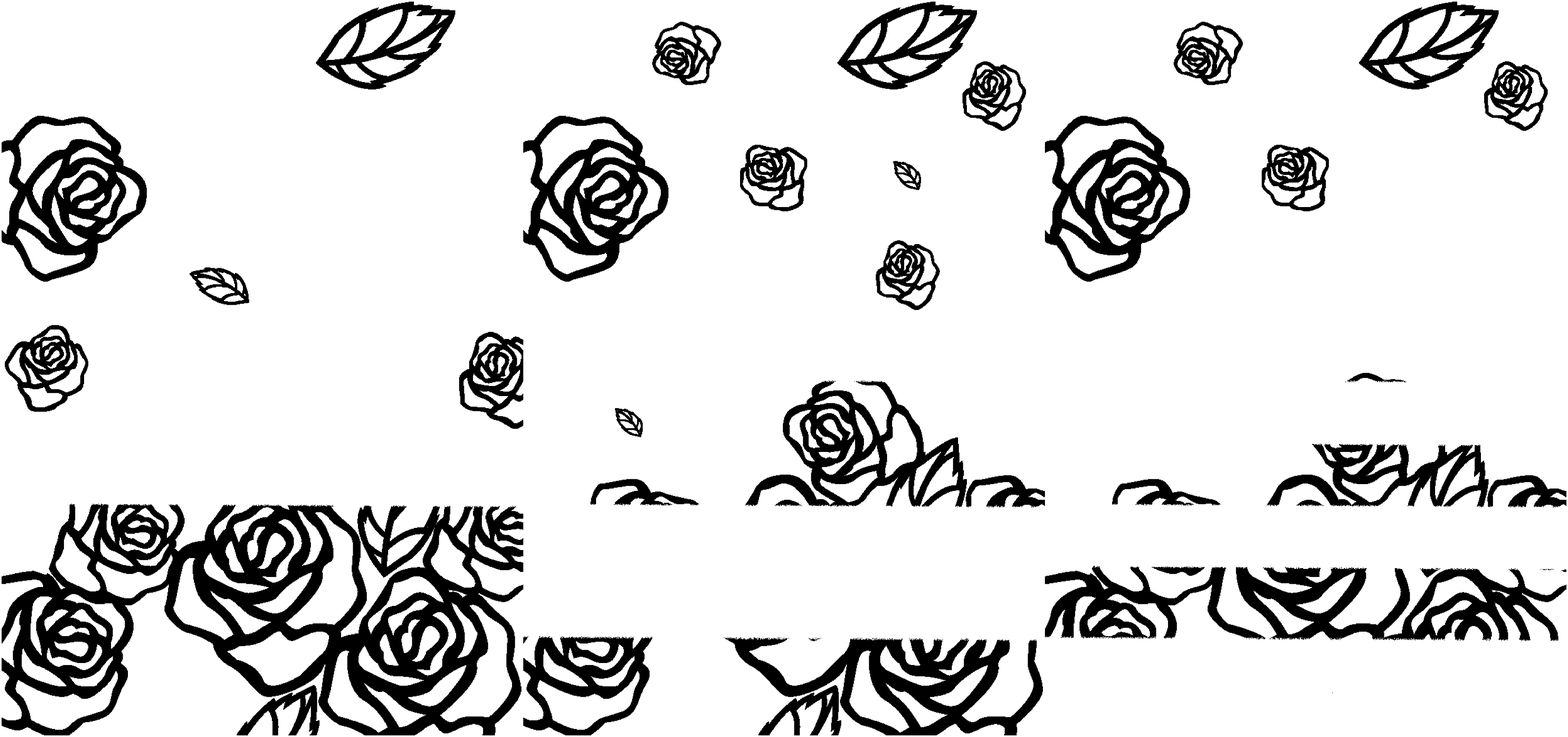}
} \\
\subfigure[Image 5]{
\includegraphics[width=0.36\linewidth]{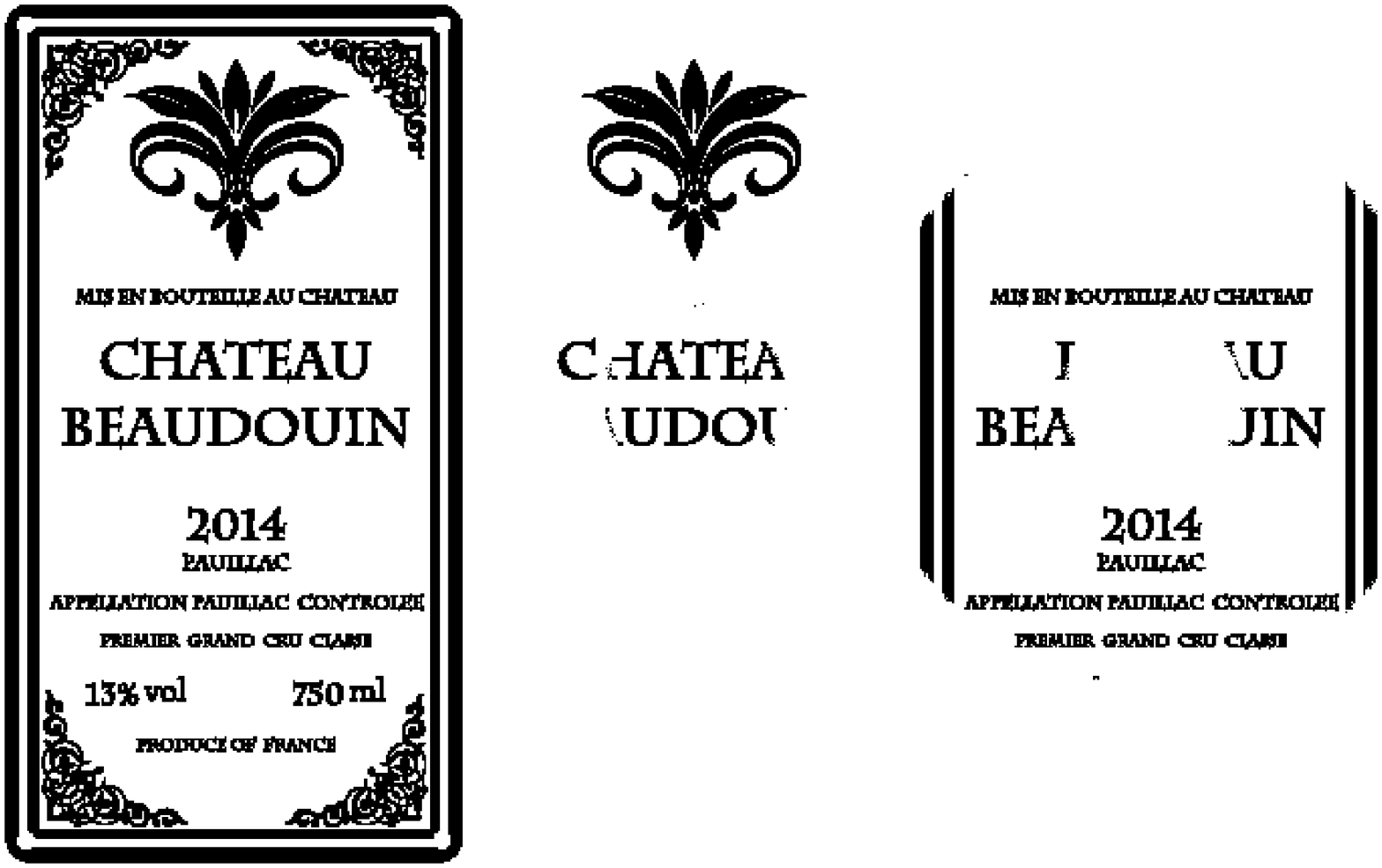}
}
\caption{3-bit special-color layers decomposed into bit planes.}
\label{fig.9}
\end{figure*}

\subsection{Data extraction procedure}
The extraction procedure is shown in Fig.~\ref{fig.5b}. 
The detailed procedure is described below.
It should be noted that $ZP$ is replaced with $LP$ in the following steps when $LP$ is used instead of $ZP$.
\begin{description}
\item[Step 1] $PP$ and $ZP$ are extracted from the LSBs of the first 16 pixels in the bottom row.
\item[Step 2] Extract the payload from the pixels $x_{PP}$ and $x_{PP+1}$/$x_{PP-1}$, where the pixel values are $PP$ and $PP+1$/$PP-1$.
The embedded bit '0' is extracted from $x_{PP}$, and '1' is extracted from $x_{PP+1}$ ($PP<ZP$) or $x_{PP-1}$ ($PP>ZP$).
\item[Step 3] All the pixels $x'$ between $PP$ and $ZP$ are shifted according to the following equation.
  \begin{equation}
       \hat{x} = \left\{ \begin{array}{ll}
         x' - 1, \quad x' \in (PP, ZP] ~~ & {\rm if} ~~ PP<ZP \\
         x' + 1, \quad x' \in [ZP, PP) ~~ & {\rm if} ~~ PP>ZP,
  \end{array} \right.
\end{equation}
where $\hat{x}$ is the retrieved pixel and should correspond to the original pixel $x$.
\item[Step 4] Decompress the compressed special-color layers, which were extracted in Step 2, by JBIG2. 
\item[Step 5] Concatenate each bit plane to restore the 3-bit special-color layer.
\item[Step 6] Restore the original of the general color layer by shifting histogram using the side information.
\end{description}

In Step 2, it is possible to obtain the side information, which contains the original LSBs of the first 16 pixels in the bottom row and the pixels that originally possess $LP$, as well as the compressed special-color layers. 
Using the side information, the original of the general color layer is restored. 
Consequently, the originals of the general color and special color layers can be completely retrieved by the data extraction process. 
As shown in Fig.~\ref{fig.1}, both the retrieved general-color and special-color layers are sent to the printer for printing with special color inks, and they are printed together.

\begin{table*}[t]
\caption{Comparison of data amount of special color layers before/after JBIG2 compression.}
\begin{center}
\begin{tabular}{|c|c|r@{\ $\times$\ }r|r|r|r|} \hline
 \multicolumn{2}{|c|}{} & \multicolumn{2}{|c|}{\begin{tabular}{c} Image size\\ $\textrm{[pixels]}$ \end{tabular}} & \begin{tabular}{c}Before compression\\ $\textrm{[bytes]}$ \end{tabular} & \begin{tabular}{c}After compression\\ $\textrm{[bytes]}$ \end{tabular} & \begin{tabular}{c}Compression ratio\\ $\textrm{[\%]}$  \end{tabular} \\ \hline
\multirow{2}{*}{Image 1} & Binary layer & 830 & 1,170 & 121,692 & 12,269 & 80.92 \\ \cline{2-7}
& 3-bit layer & 2,490 & 1,170 & 365,053 & 23,997 & 93.43 \\ \hline
\multirow{2}{*}{Image 2} & Binary layer & 830 & 1,170 & 121,692 & 2,963 & 97.57 \\ \cline{2-7}
& 3-bit layer & 2,490 & 1,170 & 365,053 & 6,778 & 98.14 \\ \hline
\multirow{2}{*}{Image 3} & Binary layer & 830 & 1,170 & 121,692 & 5,683 & 95.33 \\ \cline{2-7}
& 3-bit layer & 2,490 & 1,170 & 365,053 & 10,822 & 97.04 \\ \hline
\multirow{2}{*}{Image 4} & Binary layer & 830 & 1,170 & 121,692 & 6,756 & 94.45 \\ \cline{2-7}
& 3-bit layer & 2,490 & 1,170 & 365,053 & 12,626 & 96.54 \\ \hline
\multirow{2}{*}{Image 5} & Binary layer & 227 & 427 & 12,394 & 2,569 & 79.27 \\ \cline{2-7}
& 3-bit layer & 681 & 427 & 36,733 & 3,667 & 90.02 \\ \hline
\end{tabular}
\end{center}
\label{table2}
\end{table*}

\begin{figure*}[t]
\centering
\subfigure[Image 1]{
\includegraphics[width=0.15\linewidth]{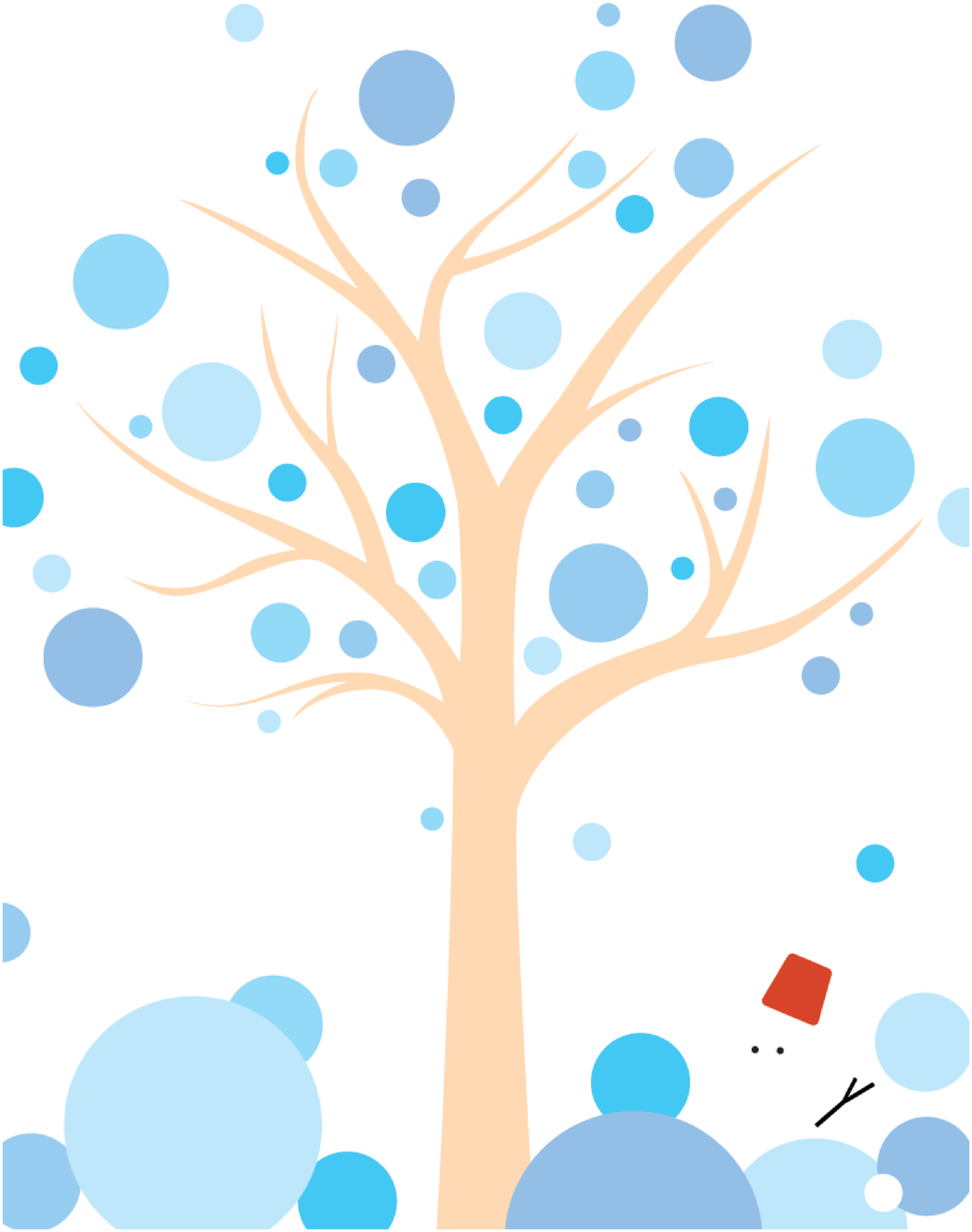}
}
\subfigure[Image 2]{
\includegraphics[width=0.15\linewidth]{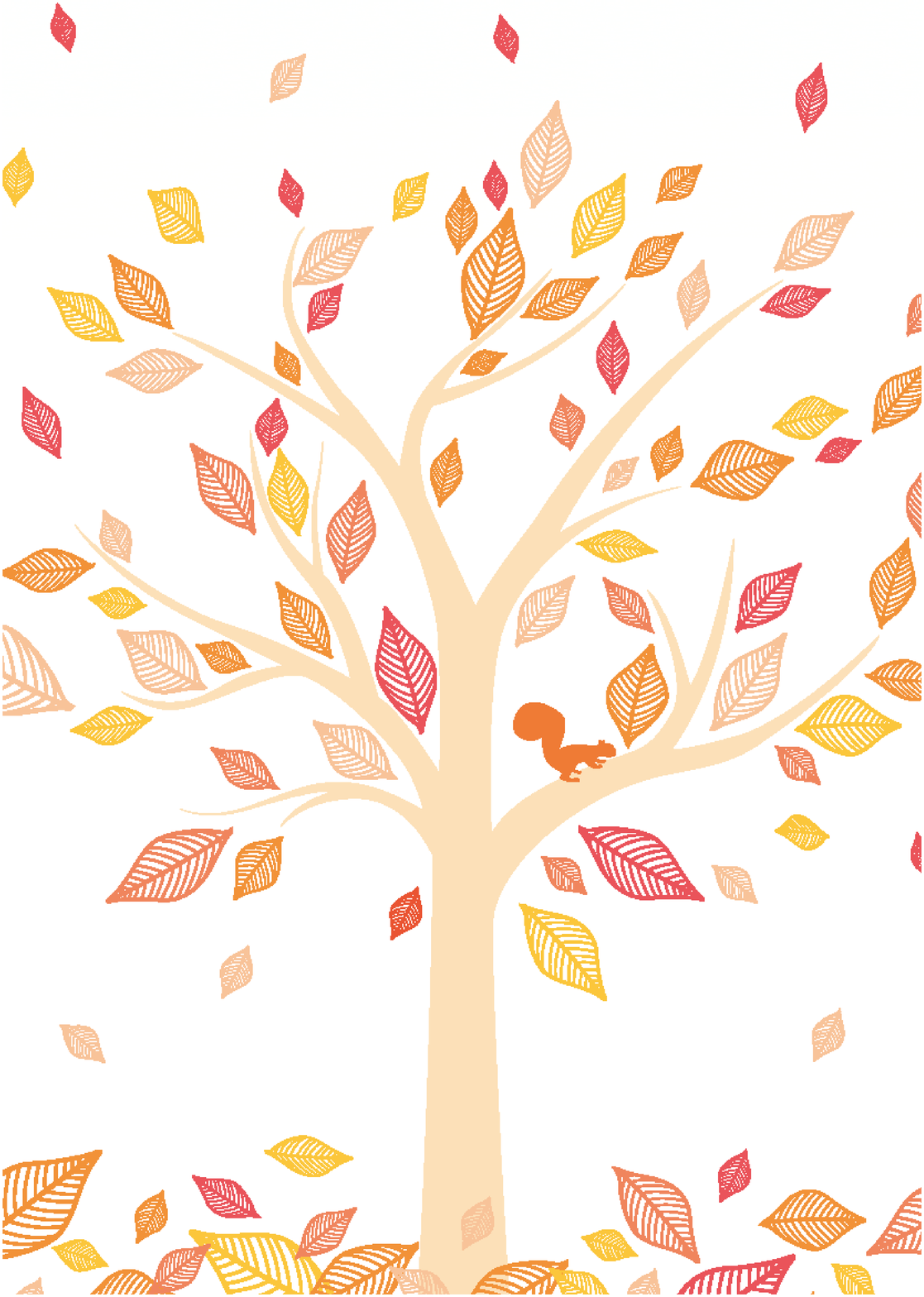}
}
\subfigure[Image 3]{
\includegraphics[width=0.15\linewidth]{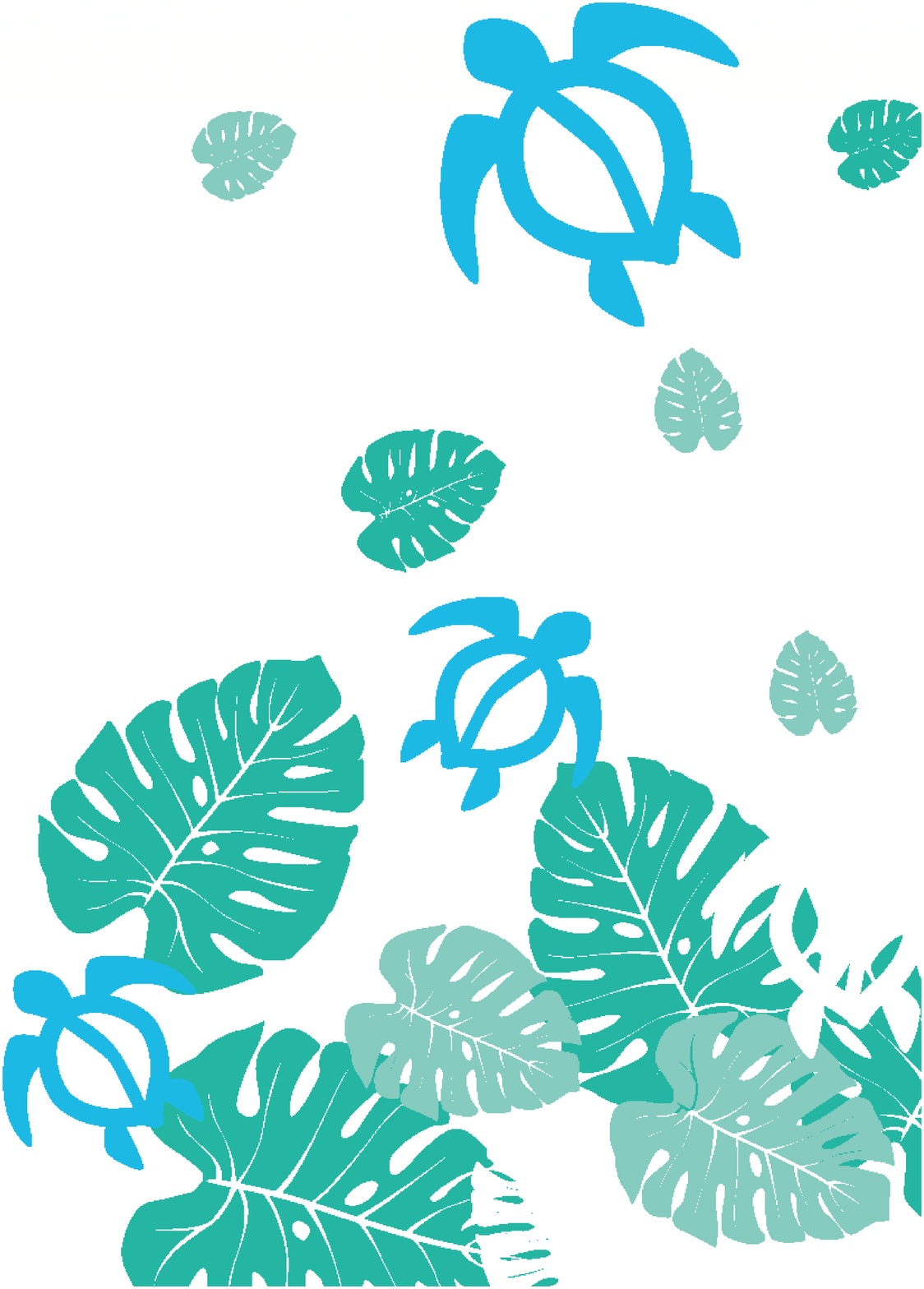}
}
\subfigure[Image 4]{
\includegraphics[width=0.15\linewidth]{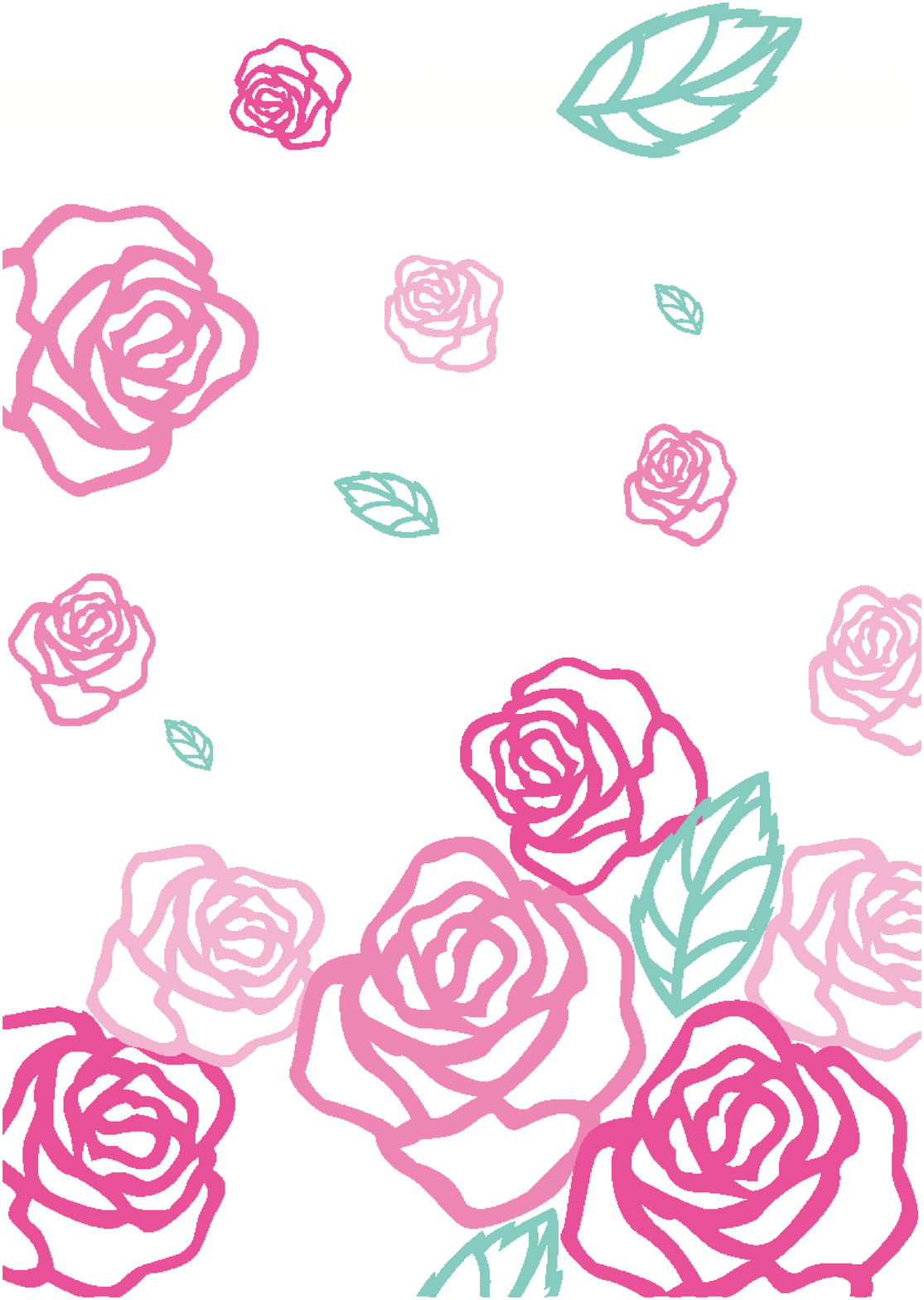}
}
\subfigure[Image 5]{
\includegraphics[width=0.12\linewidth]{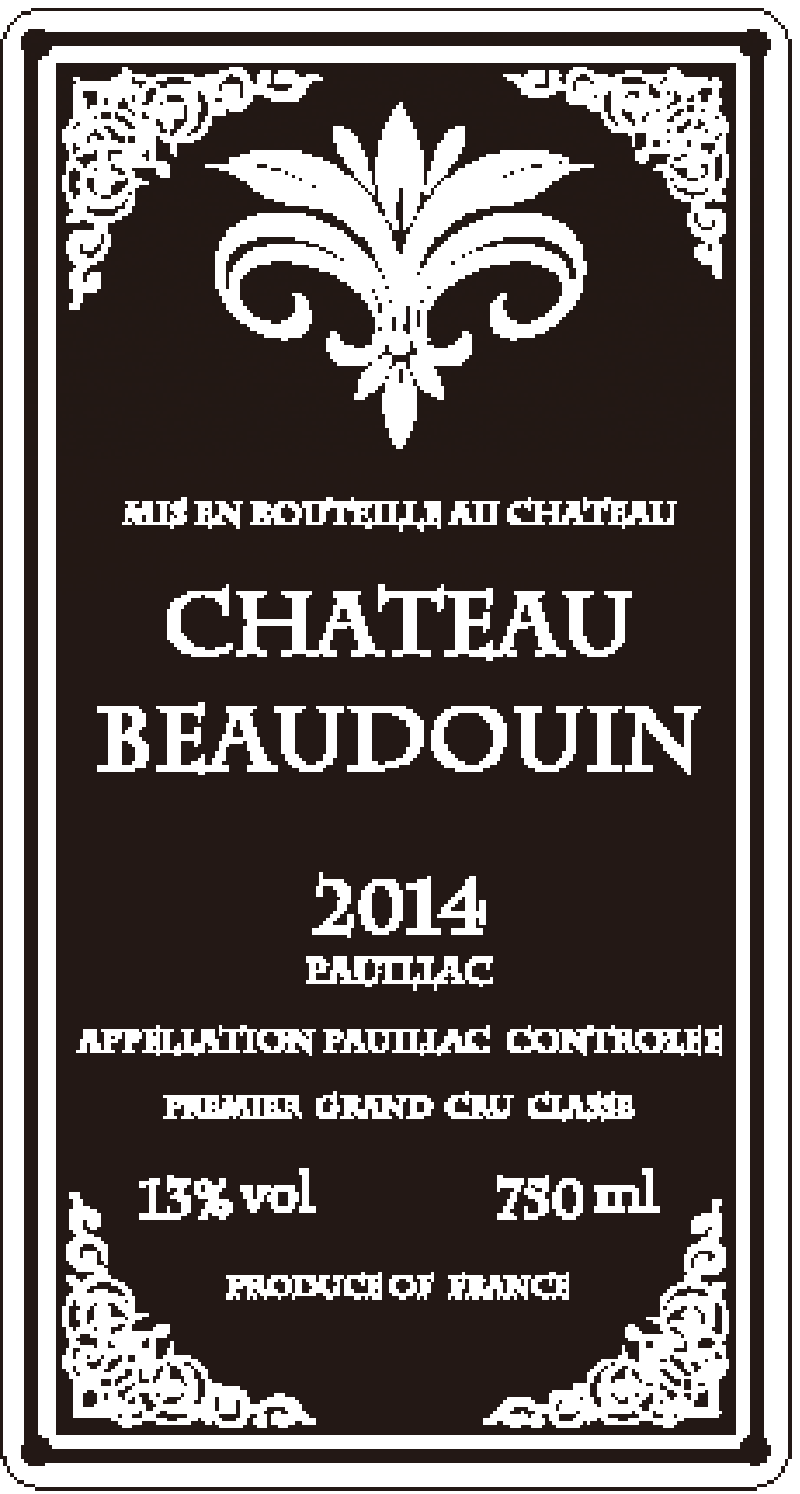}
}
\caption{Marked images.}
\label{fig.10}
\end{figure*}

\begin{table}[t]
\caption{Evaluation of marked image quality.}
\begin{center}
\begin{tabular}{|c|c|c|c|c|c|c|} \hline
& \multicolumn{2}{|c|}{Luminance} & \multicolumn{2}{|c|}{R component} & \multicolumn{2}{|c|}{B component} \\ \cline{2-7}
& PSNR [dB] & MSSIM & PSNR [dB]  & MSSIM & PSNR [dB]  & MSSIM \\ \hline
Image1 & $\approx$Inf & $\approx$1.000 & 67.36
 & 0.9992 & 62.14 & 0.9965 \\ \hline
Image2 & $\approx$Inf & $\approx$1.000 & 61.10 & 0.9996 & 58.22 & 0.9993 \\ \hline
Image3 & $\approx$Inf & $\approx$1.000 & 64.50 & 0.9998 & 61.65 & 0.9997 \\ \hline
Image4 & $\approx$Inf & $\approx$1.000 & 63.74 & 0.9998 & 61.01 & 0.9996 \\ \hline
Image5 & $\approx$Inf & $\approx$1.000 & 57.97 & 0.9995 & 56.46 & 0.9993 \\ \hline
\end{tabular}
\end{center}
\label{table3}
\end{table}

\begin{table}[t]
\caption{Marked image quality of additional 100 illustrations (B component).}
\begin{center}
\begin{tabular}{|c|c|c|c|c|} \hline
& Average & Variance & Minimum & Maximum \\ \hline
PSNR [dB] & 57.44 & 21.58 & 49.27 & 66.83 \\ \hline
MSSIM & 0.9972 & $1.877 \times 10^{-6}$ & 0.9930 & $\approx$1.000  \\ \hline
\end{tabular}
\end{center}
\label{table4}
\end{table}

\section{Experimental results and analysis}
We evaluate the quality of the marked images produced using the proposed method in terms of both objective and subjective measures.
We first used five 24-bit color illustrations, as shown in Fig.~\ref{fig.7}.
Each illustration is equivalent to the general color layer. 
The special color layers for each are designed by quantizing the Y component. 
The payload in this experiment is a pseudo-random number sequence, and the payload amount is equal to the data hiding capacity of each illustration. 

\subsection{Objective evaluation}
For objective measures, we use PSNR and SSIM\cite{ssim} in our experiment. 
PSNR is an image quality metric that calculates the similarity between two images on a pixel-by-pixel basis.
The PSNR value between the original image $X$ and reference image $Y$ is given as
\begin{equation}
PSNR(X,Y) = 10 \log_{10} \frac{MAX^2}{MSE}~[\textrm{dB}],
\end{equation}
where $MAX$ is the maximum value of the pixels in the images, namely, $MAX = 255$ in our experiment, and MSE is the mean squared error.
When an image size is $m \times n$ pixels, MSE between $X$ and $Y$ is represented by
\begin{equation}
MSE(X,Y) = \frac{1}{mn} \sum_{i=0}^{m-1} \sum_{j=0}^{n-1} (X(i,j) - Y(i,j))^2. 
\end{equation}

Meanwhile, SSIM is another image quality metric that is closer to subjective assessment than PSNR.
MSSIM is given as the average of SSIMs, which are calculated window by window.
The window determines the area for calculating each SSIM and moves pixel by pixel over the entire image.
SSIM between the window of the original image $x$ and the window of the reference image $y$, and MSSIM between $X$ and $Y$ are defined as
\begin{equation}
SSIM(x,y)=\frac{( 2 \mu_x \mu_y + C_1 ) (2 \sigma _{xy} + C_2)}{(\mu_x^2+\mu_y^2+C_1)(\sigma_x^2+\sigma_y^2+C_2)},
\label{eq:SSIM}
\end{equation}
\begin{equation}
{MSSIM(X,Y) = \frac{1}{M} \sum_{j=1}^{M} SSIM(x_j, y_j)},
\label{eq:SSIM}
\end{equation}
where $\mu$ is the average, $\sigma_x$ and $\sigma_y$ are the standard deviations, $\sigma_{xy}$ is the covariance of the pixels in each window, and $M$ is the number of windows, respectively.
In our experiment, the window size $N$ is set to 11, which is the default value. Additionally, $C_1$ and $C_2$ are estimated by
\begin{equation}
C_1=(K_1 L)^2,
\end{equation}
\begin{equation}
C_2=(K_2 L)^2,
\end{equation}
where $K_1 = 0.01$ and $K_2 = 0.03$, and $L$ is the dynamic range of the image, that is, $L = 255$. 
Note that the values of $K_1$ and $K_2$ used in this experiment are also default values, and they are most preferred in SSIM evaluation\cite{ssim}.

We embedded the compressed special-color layers, where the forms before compression are depicted in Figs.~\ref{fig.8} and \ref{fig.9}, into each general color layer. 
Table~\ref{table2} shows the comparison of the data amount of the special color layers before and after JBIG2 compression. 
It is clear that the special color layers can be highly compressed by JBIG2. 
In practice, reduction of the payload amount contributes to improve marked-image quality.

Figure~\ref{fig.10} depicts the marked images containing the compressed special-color layers.
It is difficult to subjectively perceive any distortion in the marked images.
The PSNR and MSSIM values are summarized in Table~\ref{table3}.
Here, the PSNR and MSSIM values were also derived using only the R or B component, where the payload has been embedded. 
It is clear that the marked image quality is particularly high without visible distortion.
For printing with special color inks, the general color and special color layers can be completely retrieved from the marked image through the extraction process. 

In order for the results to be statistically meaningful, we averaged on a larger set of illustrations. 
We used additional 100 illustrations from an illustration database. 
Table~\ref{table4} shows the PSNR and MSSIM values for the B component. 
In case that three color components are used to obtain the PSNR and MSSIM values for luminance, the PSNR values are particularly close to infinity and the MSSIM values are exceedingly close to one as is the case in Image 1 -- 5 shown in Table~\ref{table3}.
It was confirmed that the marked images derived in the proposed method preserve high quality.

\subsection{Subjective evaluation}
In the proposed method, the marked image is printed instead of the original of the general color layer in normal printing.
The retrieved general color and special color layers are the same as the original ones, but the marked image is slightly different from the original one.
We conducted a subjective evaluation to show that this difference is acceptable in normal printing.

\begin{figure}[t]
\begin{center}
\subfigure[RGB(0, 0, 127)]{ 
\includegraphics[width=3cm]{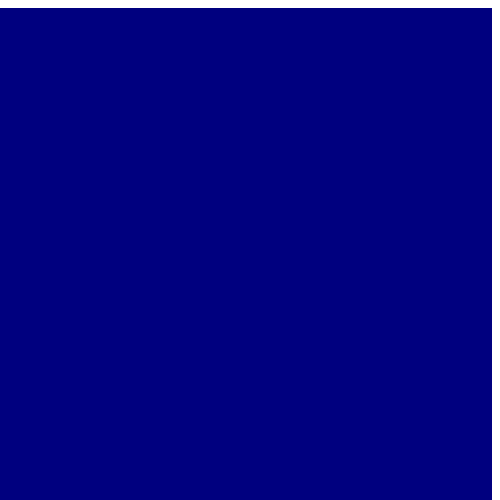}
}
\subfigure[RGB(0, 0, 128)]{
\includegraphics[width=3cm]{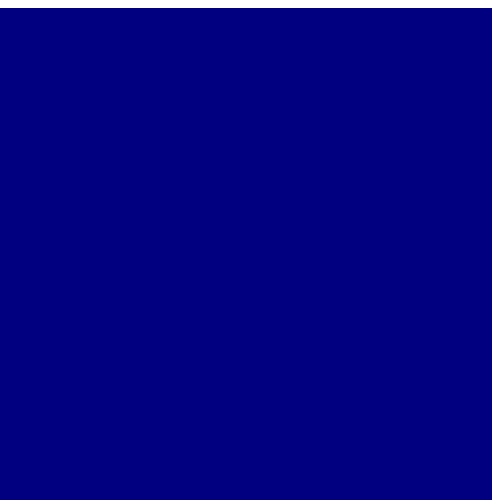}
}
\caption{Example of color patch for 2AFC.}
\label{fig.11}
\end{center}
\end{figure}

The subjective evaluations were carried out by asking participants to observe the printed color and the printed marked color.
The pixel value of the marked color is shifted by 1 from the original color.
The samples consist of three pairs: i) RGB(0, 0, 0) and RGB(0, 0, 1), ii) RGB(0, 0, 127) and RGB(0, 0, 128), and iii) RGB(0, 0, 254) and RGB(0, 0, 255). 
Figure~\ref{fig.11} illustrates the second pair, that is, RGB(0, 0, 127) and RGB(0, 0, 128). 
In the evaluation, the illuminance was approximately 2,000 lx, and the color temperature was approximately 4,000 K.

Six participants aged 20 to 25 years underwent the color discrimination experiment by 2-alternative forced choice task (2AFC).
They compared three printed colors, one of which was different from the others.
The combination of colors was presented at random, and participants picked a different color.
This was repeated ten times for each pair.
There were no participants who answered that they found differences for all color pairs correctly.
The results of the subjective evaluations indicate that the marked color is perceived as almost identical to the original color.
In addition, the marked area, which contains the payload, is much smaller than the area of the whole image.
Therefore, it is more difficult to perceive the differences, and the marked image can be printed as the general color layer in normal printing.

\section{Conclusions}
We proposed an efficient framework for compatibility between normal printing and printing with special color inks. 
In our framework, the special color layers are reversibly embedded into the general color layer after preprocessing without visible artifacts.
Consequently, the marked image can be directly printed without the extraction process in normal printing. 
In contrast, on printing with special color inks, the special color layers are extracted from the marked image, and the originals of the general color and special color layers are sent to the printer. 
Thus, the image that needs to be stored in the proposed method is only the marked image. 
This contributes to reduce the total amount of data for printing with special color inks. 
We evaluated the quality of the marked images through the experiment and confirmed that the marked images preserve high quality. 

Our future work involves an extension of the proposed method, where the target image is a natural image and the special color layer has more than 3 bits. 
We should reduce the data amount of the special color layers in preprocessing.
Additionally, we suppose not only RGB color space but also CMYK color space to achieve a more flexible printing technique.


\begin{biography}
\profile{n}{Kotoko Hiraoka}
{received her B.E. and M.E. degrees from Chiba University, Japan in 2018 and 2020, respectively. She joined Sony Semiconductor Solutions Corporation in 2020. Her research interests are image security and its applications.}
\profile{n}{Kensuke Fukumoto}
{received his B.E. and M.E. degrees from Chiba University, Japan in 2018 and 2020, respectively. His research interests are material appearance, image processing, and deep learning.}
\profile{n}{Takashi Yamazoe}
{received his B.E. in Image Engineering from Tokyo Polytechnic University in 2004, and his M.S. and Ph.D. in Global Information and Telecommunication Studies from Waseda University in 2007 and 2013, respectively. From 2017 to 2019, he attended Institute for Global Prominent Research, Chiba University, as a project researcher. In 2019, he joined Seikei University, where he is currently an Assistant Professor of Faculty of Science and Technology. His recent research interests include the material appearance of visual perception and ergonomics of newly types of mobile devices.}
\profile{n}{Norimichi Tsumura}
{received his B.E., M.E., and Dr. Eng. degrees in Applied Physics from Osaka University in 1990, 1992, and 1995, respectively. He is currently an associate professor in the Department of Information and Image Sciences, Chiba University (since February 2002).}
\profile{n}{Satoshi Kaneko}
{received his B.E. in Image Science from Chiba University and joined Mimaki Engineering Co., LTD. in 1998. He has been specialist in development of color management and image processing. Now, he is the manager of Software Department and currently researching on how to evaluate image quality for printed images.}
\profile{n}{Wataru Arai}
{received his B.E., and M.E. degrees in Electrical and Electronics Engineering from Shinshu University in 2014, and 2016, respectively. He is currently member in the Software Design Department, MIMAKI ENGINEERING CO., LTD. (since April 2016). He has involved in developing screening techniques and improved image quality.}
\profile{m}{Shoko Imaizumi}
{received her B.Eng., M.Eng., and Ph.D. degrees from Tokyo Metropolitan University, Japan in 2002, 2005, and 2011, respectively.  In 2011, she joined Chiba University, where she is currently an Associate Professor of Graduate School of Engineering. From 2003 to 2004, she was with the Ministry of Education, Culture, Sports, Science and Technology of Japan. She was a Researcher at the Industrial Research Institute of Niigata Prefecture from 2005 to 2011. Her research interests include image processing and multimedia security. Dr. Imaizumi serves as a Topic Editor for MDPI Journal of Imaging and a Director for SPIJ (Society of Photography and Imaging of Japan). She is a member of IEICE, ITE, SPIJ, APSIPA, and IEEE.}
\end{biography}

\end{document}